\documentclass[12pt,a4paper,hypertex]{article}
\usepackage{jheppub}
\allowdisplaybreaks[1]
\usepackage{bm}
\usepackage{subfigure}

\title{Dynamical AdS strings across horizons}

\author[1]{Takaaki Ishii}
\author[2]{and Keiju Murata}
\affiliation[1]{University of Colorado, 390 UCB, Boulder, CO 80309, USA}
\affiliation[2]{Keio University, 4-1-1 Hiyoshi, Yokohama 223-8521, Japan}
\emailAdd{takaaki.ishii@colorado.edu}
\emailAdd{keiju@phys-h.keio.ac.jp}

\abstract{%
We examine the nonlinear classical dynamics of a fundamental string in anti-de Sitter spacetime. The string is dual to the flux tube between an external quark-antiquark pair in $\mathcal{N}=4$ super Yang-Mills theory. We perturb the string by shaking the endpoints and compute its time evolution numerically. We find that with sufficiently strong perturbations the string continues extending and plunges into the Poincar\'e horizon. In the evolution, effective horizons are also dynamically created on the string worldsheet. The quark and antiquark are thus causally disconnected, and the string transitions to two straight strings. The forces acting on the endpoints vanish with a power law whose slope depends on the perturbations. The condition for this transition to occur is that energy injection exceeds the static energy between the quark-antiquark pair.
}


\begin{document}

\maketitle

\section{Introduction}
\label{sec:intro}
On the basis of the gauge/gravity duality~\cite{Maldacena:1997re,Gubser:1998bc,Witten:1998qj}, real time processes in strongly coupled gauge theories have been extensively studied from the viewpoint of time evolution in classical gravity. An interesting application in this direction is to understand the evolution of strongly coupled plasma observed in experiments at the Relativistic Heavy Ion Collider (RHIC) and Large Hadron Collider (LHC) from the solutions to Einstein equations in dual gravitational setups~\cite{Chesler:2008hg,Chesler:2009cy,Chesler:2010bi,Chesler:2013lia,Chesler:2015wra}. On the other hand, focusing on flavor dynamics in the gauge/gravity duality, some works computed far-from-equilibrium dynamics in D3/D7-brane systems \cite{Ishii:2014paa,Hashimoto:2014yza,Hashimoto:2014xta,Hashimoto:2014dda,Ali-Akbari:2015bha}.

In \cite{Ishii:2015wua}, we studied nonlinear dynamics under small-but-finite perturbations of the flux tube between an external quark-antiquark pair in $\mathcal{N}=4$ super Yang-Mills theory using the gauge/gravity duality. The gravity description is given by an open string hanging from the AdS boundary to the bulk \cite{Rey:1998ik,Maldacena:1998im}, and this configuration is stable under linearized perturbations \cite{Callan:1999ki,Klebanov:2006jj,Avramis:2006nv,Arias:2009me}. We computed the nonlinear time evolution of the string and found that the string exhibited turbulent behaviors that the energy is transferred to high frequency modes. The direct energy cascade continued for strings oscillating in smaller than $1+3$ dimensions, resulting in cusp formation. For string motions in full $1+4$ dimensions of the AdS$_5$, the cascade changed to an inverse energy cascade in late time and no cusp formation was found. The turbulent behaviors on probes in the gauge/gravity duality were first found in the D3/D7 system and examined \cite{Hashimoto:2014yza,Hashimoto:2014xta,Hashimoto:2014dda,Hashimoto:2015psa}. There are related AdS string works \cite{Callebaut:2015fsa,Vegh:2015ska,Vegh:2015yua} which used methods and boundary conditions different from ours.

\begin{figure}
\begin{center}
\includegraphics[scale=0.6]{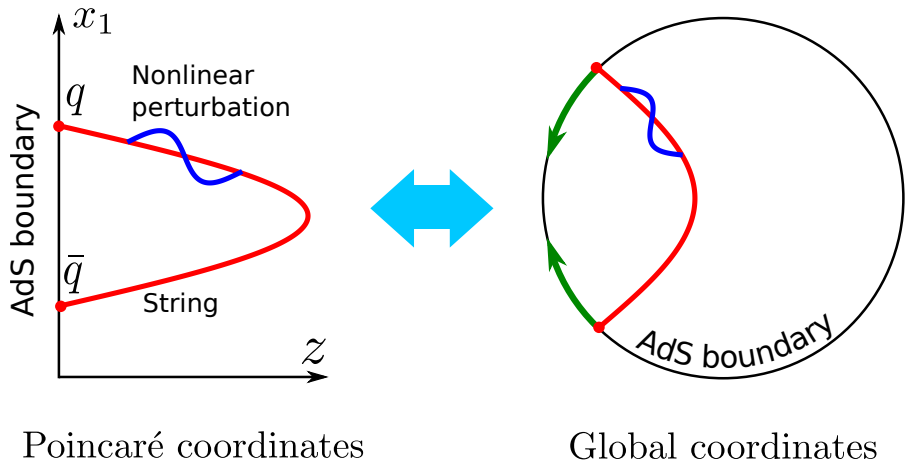}
\end{center}
\caption{%
A dynamical string in Poincar\'e and global coordinates of AdS.
Note that the string static in the Poincar\'e coordinates is mapped to a string with moving endpoints.
We solve the time evolution of the perturbed string in the global coordinates and bring it back to the Poincar\'e coordinates.
}
\label{PG_ponchi}
\end{figure}

The question we ask in this paper is what happens if the string is strongly perturbed. Initial tests using the numerical codes for \cite{Ishii:2015wua} suggested that the string would plunge into the Poincar\'e horizon. Computing in the Poincar\'e coordinates, however, had difficulties because the Poincar\'e horizon is located at a coordinate singularity. Our idea then is to make use of the global coordinates of the AdS spacetime where the Poincar\'e horizon is regular and a dynamical string can cross the Poincar\'e horizon without trouble. Since we are interested in the boundary field theory in flat space, we map the static holographic quark antiquark potential in the Poincar\'e coordinates to the global ones and solve its dynamics there. In Fig.~\ref{PG_ponchi}, we show a schematic picture of our strategy to solve the string dynamics. Note that the static string in the Poincar\'e coordinates is mapped to a nonstatic string in the global coordinates. Once computations are done, the results are brought back to the Poincar\'e patch and interpreted.

In the rest of the paper, we begin with AdS coordinates in section \ref{sec:ads}, where we introduce a parametrization convenient for our numerical work. We then prepare for computing the string dynamics in the global AdS in section \ref{sec:string}. We derive evolution equations, initial data, and boundary conditions. We also explain how to detect horizons of interest. In section \ref{sec:longi}, we show results for what we call longitudinal quenches. We first discuss a representative example and then investigate different quench parameters. In section \ref{sec:Trans}, we show results for transverse quenches where the string moves in more space dimensions than the longitudinal ones. We close the paper with a summary and several discussions in section \ref{sec:sum}. Appendices contain technical details about numerical computations.

\section{Anti-de Sitter spacetime in global and Poincar\'{e} coordinates}
\label{sec:ads}

While our interest is on a string in the Poincar\'{e} coordinates of five-dimensional anti-de Sitter (AdS$_5$) spacetime, for convenience we make use of the global coordinates of AdS$_5$. For this reason, we start from comparing these coordinates.

AdS$_5$ is given by a hypersurface embedded in six-dimensional spacetime $R^{4,2}$.
The metric of $R^{4,2}$ is
\begin{equation}
ds^2_{R^{4,2}}=-dX_0^2+dX_1^2+\cdots+dX_4^2-dX_5^2 \ .
\end{equation}
The embedding of AdS$_5$ is
\begin{equation}
-X_0^2 - X_5^2 + X_1^2 + \cdots + X_4^2 = -\ell^2 \ .
\end{equation}
where $\ell$ is a positive constant called the AdS radius.

The Poincar\'{e} coordinates cover only a part of the entire AdS spacetime.
We specify the coordinates by $t, \, z, \, x_i$ ($i=1,2,3$). The Poincar\'{e} AdS$_5$ is given by
\begin{equation}
X_0 = \frac{\ell \xi_+}{z} \ , \quad
X_i = \frac{\ell x_i}{z} \ , \quad
X_4 = \frac{\ell \xi_-}{z} \ , \quad
X_5 = \frac{\ell t}{z} \ ,
\label{Pcoord}
\end{equation}
where we defined
\begin{equation}
\xi_\pm \equiv \frac{z^2 + |\bm{x}|^2 - t^2 \pm \ell^2}{2 \ell} \ .
\end{equation}
The metric of the Poincar\'{e} AdS$_5$ is written as
\begin{equation}
ds^2=\frac{\ell^2}{z^2}(-dt^2+dz^2+d\bm{x}^2) \ ,
\end{equation}
The AdS radial coordinate $z$ takes $0 < z < \infty$. The AdS boundary is at $z=0$, and 
the null surface $z=\infty$ is called Poincar\'{e} horizon.
In these coordinates, the Poincar\'{e} horizon is at the coordinate singularity.
The boundary of the Poincar\'{e} AdS is flat $1+3$ dimensional spacetime.

For our computations with strong perturbations, it is convenient to use the global coordinates which cover the entire AdS manifold beyond the Poincar\'{e} coordinates. In particular, we take Cartesian-like coordinates $\chi_a \ (a=1,2,3,4)$ where the spatial directions are conformally flat, and the global time is denoted by $\tau$. Such global coordinates are introduced by
\begin{equation}
\frac{X_0}{X_5} = \tan \tau \ , \quad
\frac{X_a}{\ell} = \frac{2 \chi_a}{1-|\bm{\chi}|^2} \ ,
\label{Gcoord}
\end{equation}
where $\bm{\chi}=(\chi_1,\chi_2,\chi_3,\chi_4)$ satisfies $|\bm{\chi}| \le 1$. Note that both $\tau$ and $\bm{\chi}$ are dimensionless. The global AdS$_5$ metric is given by\footnote{%
Taking polar coordinates in the $\bm{\chi}$-space as $\chi_a = \tan(\theta/2) \omega_a$ where $\omega_a \ (a=1,2,3,4)$ are spherical coordinates of $S^3$, we obtain a familiar form of AdS$_5$ metric: $ds^2=\ell^2(-d\tau^2+d\theta^2+\sin^2\theta d\Omega_3^2)/\cos^2\theta$ where $d\Omega_3^2 = \sum_{a=1}^4 d\omega_a^2$ is the metric of a unit $S^3$. The polar coordinates are however ill-defined at the center of the global AdS $\theta=0$, where the $S^3$ shrinks to a point. This coordinate singularity is absent in the $\bm{\chi}$-coordinates.
}
\begin{equation}
ds^2 =-\ell^2 \left( \frac{1+|\bm{\chi}|^2}{1-|\bm{\chi}|^2} \right)^2 d\tau^2 + \frac{4 \ell^2}{(1-|\bm{\chi}|^2)^2} d\bm{\chi}^2.
\label{metric_chi}
\end{equation}
The AdS boundary locates at $|\bm{\chi}|=1$ and has the topology of $S^3$.
This metric is smooth at the AdS center $|\bm{\chi}|=0$.
Although \eqref{Gcoord} implies the presence of a closed timelike curve, we can define $\tau$ in $-\infty < \tau <\infty$ by taking the universal covering.

From \eqref{Pcoord} and \eqref{Gcoord}, we obtain the following coordinate transformation:
\begin{equation}
\begin{split}
\tau &=\tan^{-1}(t/\xi_{+})+
\begin{cases}
\pi & (t>\xi_0) \\
0 & (-\xi_0 \leq t \leq \xi_0) \\
-\pi & (t<-\xi_0)
\end{cases} \ ,\\
\chi_i &= \frac{x_i ( \sqrt{\xi_-^2 + z^2 + |\bm{x}|^2} - z )}{\xi_-^2 + |\bm{x}|^2} \quad(i=1,2,3) \ ,\\
\chi_4 &= \frac{\xi_- ( \sqrt{\xi_-^2 + z^2 + |\bm{x}|^2} - z )}{\xi_-^2 + |\bm{x}|^2} \ ,
\end{split}
\label{PtoG}
\end{equation}
where $\xi_0\equiv \sqrt{z^2+|\bm{x}|^2+\ell^2}$.
In \eqref{PtoG}, one Poincar\'{e} patch is in $-\pi \le \tau \le \pi$.
The inverse of \eqref{PtoG} is given by
\begin{align}
\frac{t}{\ell} &= \frac{(1+|\bm{\chi}|^2) \sin \tau}{(1+|\bm{\chi}|^2)\cos \tau - 2 \chi_4}\ , \label{GtoP_t}\\
\frac{z}{\ell} &= \frac{1-|\bm{\chi}|^2}{(1+|\bm{\chi}|^2) \cos \tau - 2 \chi_4} \ , \label{GtoP_z}\\
\frac{x_i}{\ell} &= \frac{2\chi_i}{(1+|\bm{\chi}|^2) \cos \tau - 2 \chi_4}\quad (i=1,2,3)\ . \label{GtoP_x}
\end{align}
From \eqref{GtoP_z}, we find that the denominator must be zero at the the Poincar\'{e} horizon.
The equation of the Poincar\'{e} horizon in the global coordinates hence becomes 
\begin{equation}
\sum_{i=1}^{3} \chi_i^2 +\left(\chi_4-\frac{1}{\cos \tau}\right)^2 = \tan^2\tau\ .
\label{poincare_horizon_sphere}
\end{equation}
This represents a $S^3$ with the center $(\chi_i,\chi_4)=(0,1/\cos\tau)$ and the radius $|\tan \tau|$.
The Poincar\'{e} horizon is given by the cross section of this $S^3$ cut off by the unit $S^3$ of the AdS boundary $|\bm{\chi}|=1$. 
Moreover, if $x_i$ is finite on the Poincar\'{e} horizon, the numerator in \eqref{GtoP_x} also has to be zero: $\chi_i=0$. 
Therefore, the part of the Poincare horizon at finite $x_i$ is mapped to a point given by $(\chi_i,\chi_4)=(0,(1-|\sin \tau|)/\cos \tau)$.

\begin{figure}
\begin{center}
\includegraphics[scale=0.55]{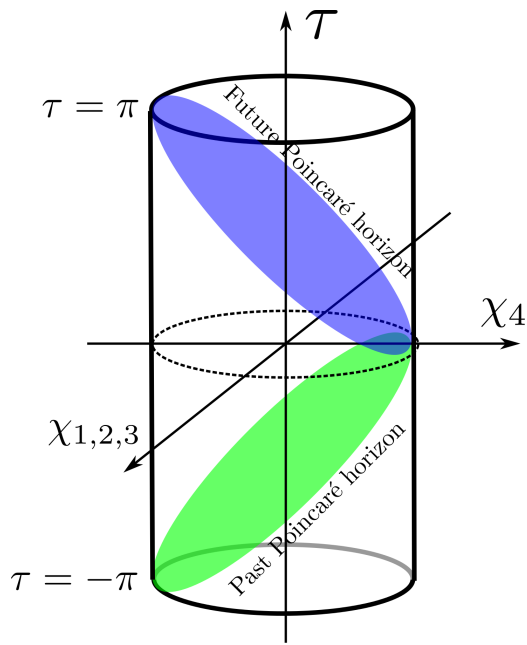}
\end{center}
\caption{%
A schematic picture of AdS$_5$ spacetime.
The inside of the cylinder is the global AdS$_5$ spacetime and its surface is the AdS boundary.
Future and past Poincar\'{e} horizons are shown by inclined surfaces in the cylinder.
The region surrounded by these surfaces corresponds to the Poincar\'{e} patch.
}
\label{gp_ponchi}
\end{figure}

In Fig.~\ref{gp_ponchi}, we show a schematic picture of AdS$_5$ spacetime in the global patch. When $\tau=\pm \pi$, the Poincar\'{e} patch shrinks to a point $(\chi_i,\chi_4)=(0,-1)$. Hence, the dynamics confined to a Poincar\'{e} patch becomes singular at that time, unless it jumps over the Poincare horizon before that time. 

\section{String dynamics in global AdS$_5$}
\label{sec:string}

We consider strong perturbations of the string in AdS$_5$. For this purpose, it is suitable to use the global coordinates~(\ref{metric_chi}) which is regular in the entire spacetime, rather than the Poincar\'{e} coordinates. Here, we describe the formulation to solve the string dynamics in the global AdS$_5$ numerically. We basically follow the method developed in \cite{Ishii:2014paa,Ishii:2015wua}.

\subsection{Evolution equations}

We make use of double null coordinates on the string worldsheet.
When the worldsheet coordinates are denoted by $(u,v)$, the string in the target space is parametrized as
\begin{equation}
 \tau=\tau(u,v)\ ,\quad \bm{\chi}=\bm{\chi}(u,v)\ .
\end{equation}
Substituting them into Eq.~(\ref{metric_chi}), 
we obtain the components of the induced metric as
\begin{align}
\gamma_{uu} &= \frac{\ell^2}{(1-|\bm{\chi}|^2)^2} \left( -(1+|\bm{\chi}|^2)^2 \, \tau_{,u}^2 + 4 |\bm{\chi}_{,u}|^2 \right)\ , \nonumber \\
\gamma_{vv} &= \frac{\ell^2}{(1-|\bm{\chi}|^2)^2} \left( -(1+|\bm{\chi}|^2)^2 \, \tau_{,v}^2 + 4 |\bm{\chi}_{,u}|^2 \right)\ , \nonumber \\
\gamma_{uv} &= \frac{\ell^2}{(1-|\bm{\chi}|^2)^2} \left( -(1+|\bm{\chi}|^2)^2 \, \tau_{,u} \tau_{,v} + 4 \bm{\chi}_{,u} \cdot \bm{\chi}_{,v} \right)\ ,
\label{gammas}
\end{align}
Using the reparametrization freedom of the worldsheet
coordinates, we impose the double null condition on the induced metric as
\begin{equation}
C_1\equiv\gamma_{uu}=0\ ,\qquad
C_2\equiv\gamma_{vv}=0\ .
\label{CON}
\end{equation}
From the double null conditions, we obtain
\begin{align}
\tau_{,u} = \frac{2 |\bm{\chi}_{,u}|}{1+|\bm{\chi}|^2} \ , \quad
\tau_{,v} = \frac{2 |\bm{\chi}_{,v}|}{1+|\bm{\chi}|^2} \ ,
\label{constraintsuv_chi}
\end{align}
where we took the positive signature regarding $\partial_u$ and $\partial_v$ as future directed null vectors.
In the double null coordinates, the Nambu-Goto action is written as
\begin{equation}
\begin{split}
S&
=-\frac{1}{2\pi\alpha'}\int dudv \sqrt{\gamma_{uv}^2-\gamma_{uu}\gamma_{vv}}
=\frac{1}{2\pi\alpha'}\int dudv \gamma_{uv}\\
&=\frac{\sqrt{\lambda}}{2\pi}\int dudv\,\frac{1}{(1-|\bm{\chi}|^2)^2} \left( -(1+|\bm{\chi}|^2)^2 \, \tau_{,u} \tau_{,v} + 4 \bm{\chi}_{,u} \cdot \bm{\chi}_{,v} \right)\ ,
\end{split}
\label{action1}
\end{equation}
where in the second equality we used the double null conditions~(\ref{CON}) and $\gamma_{uv}<0$.
The {}'t Hooft coupling is defined by $\lambda \equiv \ell^4/\alpha'{}^2$.
The string evolution equations are given by
\begin{align}
&\tau_{,uv} = -\frac{8}{(1-|\bm{\chi}|^2)(1+|\bm{\chi}|^2)^2} \left(|\bm{\chi}_{,u}| \bm{\chi}_{,v} + \bm{\chi}_{,u} |\bm{\chi}_{,v}| \right)\cdot \bm{\chi} \ , \nonumber \\
&\bm{\chi}_{,uv} = -\frac{2}{1-|\bm{\chi}|^4} \Big[ 2|\bm{\chi}_{,u}| |\bm{\chi}_{,v}| \bm{\chi}\nonumber \\
 &\hspace*{15ex} + (1+|\bm{\chi}|^2)\big\{ (\bm{\chi} \cdot \bm{\chi}_{,v})\bm{\chi}_{,u}
 + ( \bm{\chi}\cdot\bm{\chi}_{,u} ) \bm{\chi}_{,v} - (\bm{\chi}_{,u} \cdot \bm{\chi}_{,v})\bm{\chi} \big\} \Big] \ ,
\label{evol}
\end{align}
where in the right hand sides,
we eliminated $\tau_{,u}$ and $\tau_{,v}$ by using Eq.~(\ref{constraintsuv_chi}).
This process is very important for stabilizing numerical calculations.

\subsection{Poincar\'e static string in the global patch}

As the initial configuration, we consider a static string hanging in the Poincar\'e AdS from the boundary.
Its endpoints correspond to a pair of quark and antiquark.
We locate the endpoints at $x_1=\pm L/2$ and $x_2=x_3=0$ where $L$ is the separation.
We call the one with $x_1=L/2$ the quark endpoint.
The static solution parametrized by the double null coordinates $(u,v)$ was obtained in \cite{Ishii:2015wua} as
\begin{equation}
 t = z_0 \, (u+v)\ ,\quad z = z_0\, f(u-v)\ ,\quad x_1 = z_0\, g(u-v)\ , \quad x_2=x_3=0\ ,
\label{Pstatic}
\end{equation}
where $z_0\equiv L/(2\Gamma_0)$ and $\Gamma_0 \equiv \sqrt{2}\pi^{3/2}/\Gamma(1/4)^2 \simeq 0.599$.
The constant $z_0$ represents the maximum value of the $z$-coordinate that the string reaches.
The functions $f$ and $g$ are defined as
\begin{equation}
f(\phi) \equiv \textrm{sn}(\phi;i)\ ,\quad
g(\phi) \equiv-\int^\phi_{\beta_0/2}d\phi'\, f(\phi')^2\ ,
\label{fgdef}
\end{equation}
where 
$\textrm{sn}(x;k)$ is the Jacobi elliptic function.\footnote{%
The Jacobi elliptic function $\textrm{sn}(x;k)$ is the inverse function of 
the incomplete elliptic integral of the first kind defined as
$F(x;k)=\int^x_0 dt (1-t^2)^{-1/2}(1-k^2t^2)^{-1/2}$.
}
The constant $\beta_0\equiv \pi/(2\Gamma_0)\simeq2.622$ is the minimum positive root of $\textrm{sn}(x;i)$,
and in \eqref{Pstatic} we have used the worldsheet reparametrization degrees of freedom to locate the two boundaries of the open string worldsheet at $u=v$ and $u=v+\beta_0$.

\begin{figure}[t]
\centering
\includegraphics[scale=0.5]{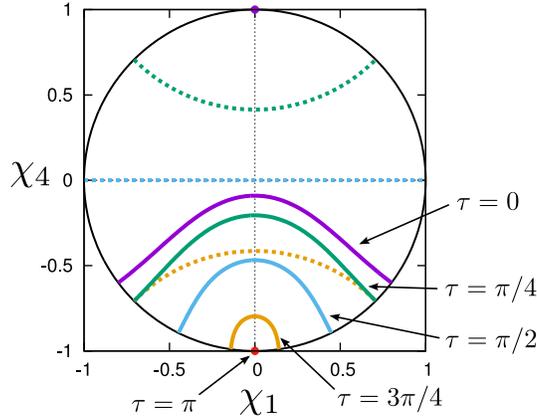}
\caption{Mapping of the Poincar\'{e} static hanging string to the global coordinates when $\ell/L=1$. The colored lines correspond to the string configurations when $\tau=0, \, \pi/4, \, \pi/2, \, 3\pi/4$ from the top, and the red dot at $(\chi_1,\chi_4)=(0,-1)$ is that for $\tau=\pi$. The three dashed lines as well as the dots at $(\chi_1,\chi_4)=(0,\pm1)$ are the Poincare horizon at each corresponding time. Their ordering from the top is the same as that of the string.}
\label{fig:initF1}
\end{figure}

We map the Poincar\'e patch's static solution (\ref{Pstatic}) to the global coordinates.
This is done by substituting it into Eq.~(\ref{PtoG}).
It is straightforward to check that the mapped solution satisfy the evolution equations~(\ref{evol}) and the constraints~(\ref{CON}). The solution static in the Poincar\'{e} patch~(\ref{Pstatic}) becomes a $\tau$-dependent solution in the global coordinates.
In Fig.~\ref{fig:initF1}, we show the string configurations for several $\tau$-slices. Note that $\chi_2=\chi_3=0$ for the static solution, and hence Fig.~\ref{fig:initF1} corresponds to the spatial part of an AdS$_3$ slice in the AdS$_5$. The string shrinks as $\tau$ approaches $\pi$ as the Poincar\'{e} horizon surrounding the string also does. Eventually, both the static string and the Poincar\'e horizon collapse to a point at $\tau=\pi$. The string configurations are symmetric under $\tau \to -\tau$, and thus the string expands as $\tau$ increases for $-\pi \le \tau \le 0$.

In the Poincar\'e coordinates, the AdS radius $\ell$ does not appear in the equations and solution of the string. However, the transformation \eqref{PtoG} involves $\ell$, and therefore the string configurations in the global coordinates look differently depending on $\ell/L$. For instance, Fig.~\ref{fig:initF1} is drawn when $\ell/L=1$. Nevertheless, once the string dynamics computed in the global coordinates is transformed back to the Poincar\'e patch, the dependence on $\ell$ disappears. We can hence use arbitrary values of $\ell/L$ convenient for numerical computations.

For the time evolution, we take $v=0$ as the initial surface and use the Poincar\'{e} static solution mapped to the global coordinates as the initial data. Nontrivial string dynamics is induced by changing the boundary conditions at the string endpoints in time.

\subsection{Boundary conditions}

In the Poincar\'{e} patch, the two string endpoints are denoted by $\bm{x}=\bm{x}_q(t)$ and 
$\bm{x}_{\bar{q}}(t)$, 
corresponding to the locations of the quark and antiquark, respectively.
We induce dynamics on the string by changing their positions.
Since we are interested in the dynamics in the Poincar\'e patch,
we specify the patterns of endpoint motion as functions there.
In solving the time evolution in the global patch, the quark positions are translated to
the global coordinates' boundary through Eq.~(\ref{PtoG}).

\begin{figure}
\begin{center}
\includegraphics[scale=0.7]{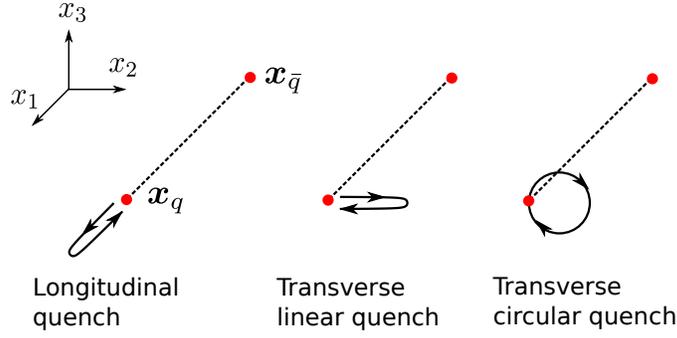}
\end{center}
\caption{%
Schematic pictures of the quenches.
}
\label{quench_ponchi}
\end{figure}

To introduce perturbations on the string, we move the quark endpoint for a time duration $\Delta t$ with amplitude $\epsilon$.
In particular, we consider the following three kinds of ``quenches'':
\begin{enumerate}
\renewcommand{\labelenumi}{\textbf{(\roman{enumi})}}
\item \textbf{Longitudinal quench}:
\begin{equation}
 \bm{x}_q(t)=\left(\frac{L}{2}+\epsilon L\alpha(t),0,0\right)\ .
\label{quench_longi}
\end{equation}
\item \textbf{Transverse linear quench}:
\begin{equation}
 \bm{x}_q(t)=\left(\frac{L}{2},\epsilon L \alpha(t),0\right)\ .
\label{quench_tranlin}
\end{equation}
\item \textbf{Transverse circular quench}:
\begin{equation}
 \bm{x}_q(t)=\left(\frac{L}{2},\epsilon L \alpha(t),\pm\epsilon L \sqrt{\alpha(t)(1-\alpha(t))}\right)\ ,
\label{quench_trancirc}
\end{equation}
where the upper and lower signs are taken for $t\leq \Delta t/2$ and $t> \Delta t/2$, respectively.
\end{enumerate}
The other endpoint is fixed at the original position: $\bm{x}_{\bar{q}}(t)=(-L/2,0,0)$. 
The function $\alpha(t)$ is a compactly supported $C^\infty$ function defined by
\begin{equation}
\alpha(t)=
\begin{cases}
\exp\left[2\left(\frac{\Delta t}{t-\Delta t}-\frac{\Delta t}{t}+4
\right)\right]\qquad &(0<t<\Delta t)\\
0\qquad &(\textrm{else})
\end{cases} \ .
\end{equation}
This function has the maximum value $1$ at $t=\Delta t/2$.
In Fig.~\ref{quench_ponchi}, we show schematic pictures of the quenches (i)-(iii).
For later convenience, we introduce the velocity and the Lorentz factor of the quark as
\begin{equation}
\bm{v}(t)\equiv \frac{d\bm{x}_q(t)}{dt}\ ,\quad
\gamma(t)\equiv\frac{1}{\sqrt{1-\bm{v}^2(t)}}\ .
\label{vgamma}
\end{equation}

These quench patterns are the same as those considered in \cite{Ishii:2015wua} and chosen to represent typical string motions, particularly with different dimensionality. For quenches (i) and (ii),
the motion of the string is restricted in (2+1)- and (3+1)-dimensions, respectively.
On the other hand, by the quench (iii), 
fluctuations along both $x_2$- and $x_3$-directions are induced, and the string moves in all (4+1)-dimensions.
The significant difference from \cite{Ishii:2015wua} in this paper is the magnitude of $\epsilon$, which we choose much bigger.

Using the residual coordinate freedoms $u=u(\bar{u})$ and $v=v(\bar{v})$, 
we fix the locations of the string endpoints on the worldsheet to $u=v$ and $u=v+\beta_0$. 
Letting $\bm{\chi}_q(\tau)$ and $\bm{\chi}_{\bar{q}}(\tau)$ denote the trajectories of the endpoints in the global patch, 
we write the boundary conditions for $\bm{\chi}(u,v)$ as
\begin{equation}
\bm{\chi}|_{u=v}=\bm{\chi}_q(\tau)\ ,\quad
\bm{\chi}|_{u=v+\beta_0}=\bm{\chi}_{\bar{q}}(\tau)\ .
\label{bdrycond_chi}
\end{equation}
The functional form of $\bm{\chi}(\tau)$ is found from $(\tau,\bm{\chi})$ through \eqref{PtoG} where the right hand sides are determined by the quark and antiquark positions in the Poincar\'e patch's boundary. While the endpoints are quenched, it is hard to find analytic expressions of $\bm{\chi}(\tau)$. It is practical to find the relation between $\tau$ and $\bm{\chi}$ numerically at each time.

Outside the quench period, however, simple expressions for \eqref{bdrycond_chi} can be obtained.
When $\bm{x}=(\tilde{x},0,0)$ is constant, we can eliminate $t$ and $z$ from \eqref{PtoG} and obtain
\begin{equation}
\begin{split}
\chi_1 |_\mathrm{bdry} &=\frac{\ell \tilde{x} \cos \tau + \tilde{x} \sqrt{\tilde{x}^2 \sin^2 \! \tau + \ell^2}}{\tilde{x}^2 + \ell^2} \ , \\
\chi_4 |_\mathrm{bdry} &=\frac{\tilde{x}^2 \cos \tau - \ell \sqrt{\tilde{x}^2 \sin^2 \! \tau + \ell^2}}{\tilde{x}^2 + \ell^2} \ ,
\end{split}
\label{boundary_static_chi_evo}
\end{equation}
with $\chi_2 = \chi_3 = 0$. In our case, $\tilde{x}=\pm L/2$.

At the boundaries, we evolve $\tau(u,v)$ by satisfying the boundary conditions \eqref{bdrycond_chi}.
At $|\bm{\chi}|=1$, the constraint equations \eqref{constraintsuv_chi} reduce to
\begin{align}
\tau_{,u} = |\bm{\chi}_{,u}| \ , \quad \tau_{,v} = |\bm{\chi}_{,v}|\ .
\label{bdryTevo_chi}
\end{align}
Solving these equations, we determine the consistent evolution of $(\tau,\bm{\chi})$.

To solve the time evolution, we use the numerical method explained in \cite{Ishii:2015wua}. (See its Appendix A for details.) We introduce a grid with spacing $\Delta u = \Delta v =h$ on the $(u,v)$-coordinates and then discretize the equations of motion with second-order finite differentials. How to discretize the boundary evolution \eqref{bdryTevo_chi} for this global AdS case is described in Appendix~\ref{sec:disc}. In Appendix~\ref{sec:error}, numerical errors are evaluated.

\subsection{String dynamics and horizons}
Solving the evolution equations~(\ref{evol}), we obtain $\tau(u,v)$ and $\bm{\chi}(u,v)$ as functions on the worldsheet. We compare the solution with horizons relevant to our string dynamics.

\subsubsection*{Poincar\'e horizon}
One of this paper's interests is to see whether the string can reach the Poincar\'e horizon. In the global coordinates, the target space metric is regular at the Poincar\'e horizon. Hence there is no obligation for the dynamical string to cross the surface. To check if this occurs, we compare the solution with \eqref{poincare_horizon_sphere}. On the worldsheet, the locations crossing the Poincar\'e horizon, if exist, are continuous and form a spacelike curve connecting the two boundary points with $\tau=\pi$.

Whether the Poincar\'e horizon is crossed or not also divides the fate of the string evolution. If the string does not cross the horizon, it shrinks to one point $(\chi_1,\chi_2,\chi_3,\chi_4)=(0,0,0,-1)$ at $\tau=\pi$. On the worldsheet point of view, the fields $\tau(u,v)$ and $\bm{\chi}(u,v)$ converge to the limiting values as $(u,v) \to \infty$. However, at $\tau=\pi$, denominators in the right hand sides of the evolution equations~(\ref{evol}) become close to zero, and numerical calculations break down as $\tau=\pi$ is approached. This fate is different if the string cross the Poincar\'e horizon by strong quenches. In this case, the string extends in the global AdS at $\tau=\pi$. Then \eqref{evol} is regular, and the numerical calculations can continue to $\tau>\pi$. In Fig.~\ref{effhor}, we show the string worldsheet for such a case schematically.

\subsubsection*{Worldsheet effective horizons}
\begin{figure}[t]
\centering
\includegraphics[scale=0.5]{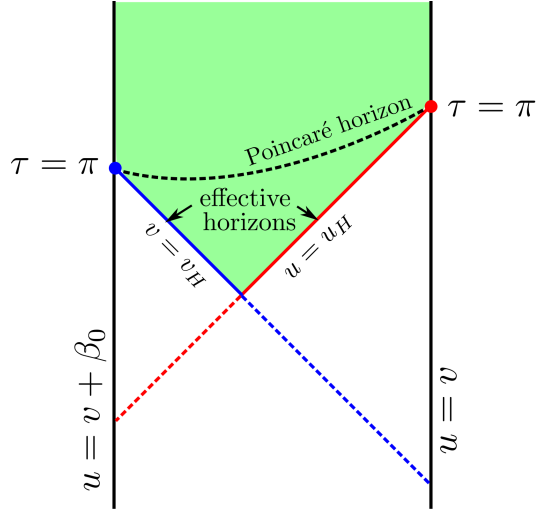}
\caption{Poincar\'e and effective horizons on the worldsheet.}
\label{effhor}
\end{figure}

The part of the string worldsheet that have crossed the Poincar\'e horizon, of course, is not visible from the boundary in a Poincar\'e patch. Before the Poincar\'e horizon, the string worldsheet has dynamically created effective horizons from the inside of which no signal reaches the AdS boundary in the Poincar\'e patch. We can define the effective horizons as the boundary of the causal past of the string endpoints at $\tau=\pi$. In Fig.~\ref{effhor}, the effective horizons are drawn with blue and red lines, which are respectively at constant $v$ and $u$. We also drew the Poincar\'e horizon with a black dashed curve. The Poincar\'e horizon is hidden in the effective horizons.
Once the string plunges into the Poincar\'e horizon,
the quark and antiquark are disconnected informationally in the Poincar\'e patch.
For this reason, we will sometimes refer to the plunge of the string into the Poincar\'e horizon as the {\textit{disconnection}} of the string

Practically, we can identify the effective horizons in the numerical solutions as follows.
We monitor $\tau(u,v)$ at each boundary and find $u=u_H$ and $v=v_H$ such that 
$\tau(u_H,u_H)=\pi$ and $\tau(v_H+\beta_0,v_H)=\pi$, respectively.
(These are shown by red and blue points in Fig.~\ref{effhor}.)
Then, the effective horizon is given by
$\mathcal{H}=\{(u,v)|u=u_H, v\geq v_H\}\bigcup\{(u,v)|v=v_H, u\geq u_H\}$.
This specify the region that cannot be seen from both endpoints. This part is shaded with green in Fig.~\ref{effhor}. Note that each of the triangular region surrounded by the dashed and solid lines and a boundary can be seen only from one endpoint.

\section{Longitudinal quench}
\label{sec:longi}

We show results of string dynamics. We start from the longitudinal quench \eqref{quench_longi}, which exhibits typical phenomena. We will also discuss the transverse quenches \eqref{quench_tranlin} and \eqref{quench_trancirc} in the following section, emphasizing similarities and differences.

\subsection{String dynamics}

\begin{figure}[t]
\centering
\includegraphics[scale=0.55]{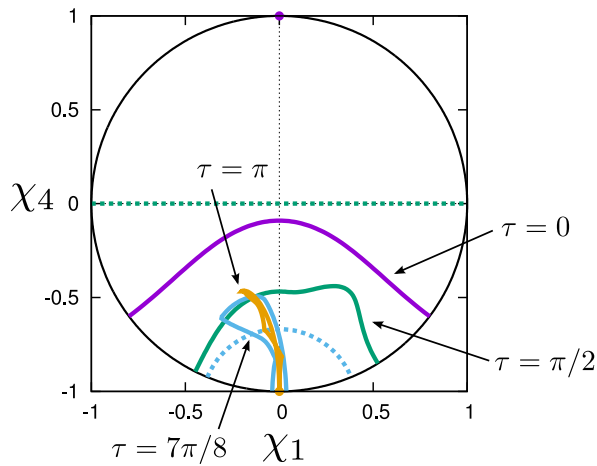}
\caption{Snapshots of the string dynamics in the global patch for the longitudinal quench with $\epsilon=0.15$ and $\Delta t/L=2$ for $\ell/L=1$. The real lines correspond to the string, and the dashed lines the Poincar\'e horizon.}
\label{LLwhole}
\end{figure}

As a representative example, we consider the longitudinal quench with $\epsilon=0.15$ and $\Delta t/L=2$.
In Fig.~\ref{LLwhole}, we show snapshots of the string in the global patch for $\ell/L=1$. Starting from a static hanging string configuration at $\tau=0$, we observe that a wave is induced on the string by the quench ($\tau=\pi/2$). As $\tau=\pi$ approaches, we find that the string extends beyond the Poincar\'e horizon ($\tau=7\pi/8$), and when $\tau=\pi$ the string keeps a finite length, in contrast to the shrinking static string in Fig.~\ref{fig:initF1} (and strings with small perturbations as well).
We thus find that this parameter choice corresponds to a case where the dynamical string crosses the Poincar\'e horizon.

\begin{figure}
\centering
\subfigure[$2\leq \tau \leq 2.3$]
{\includegraphics[scale=0.45]{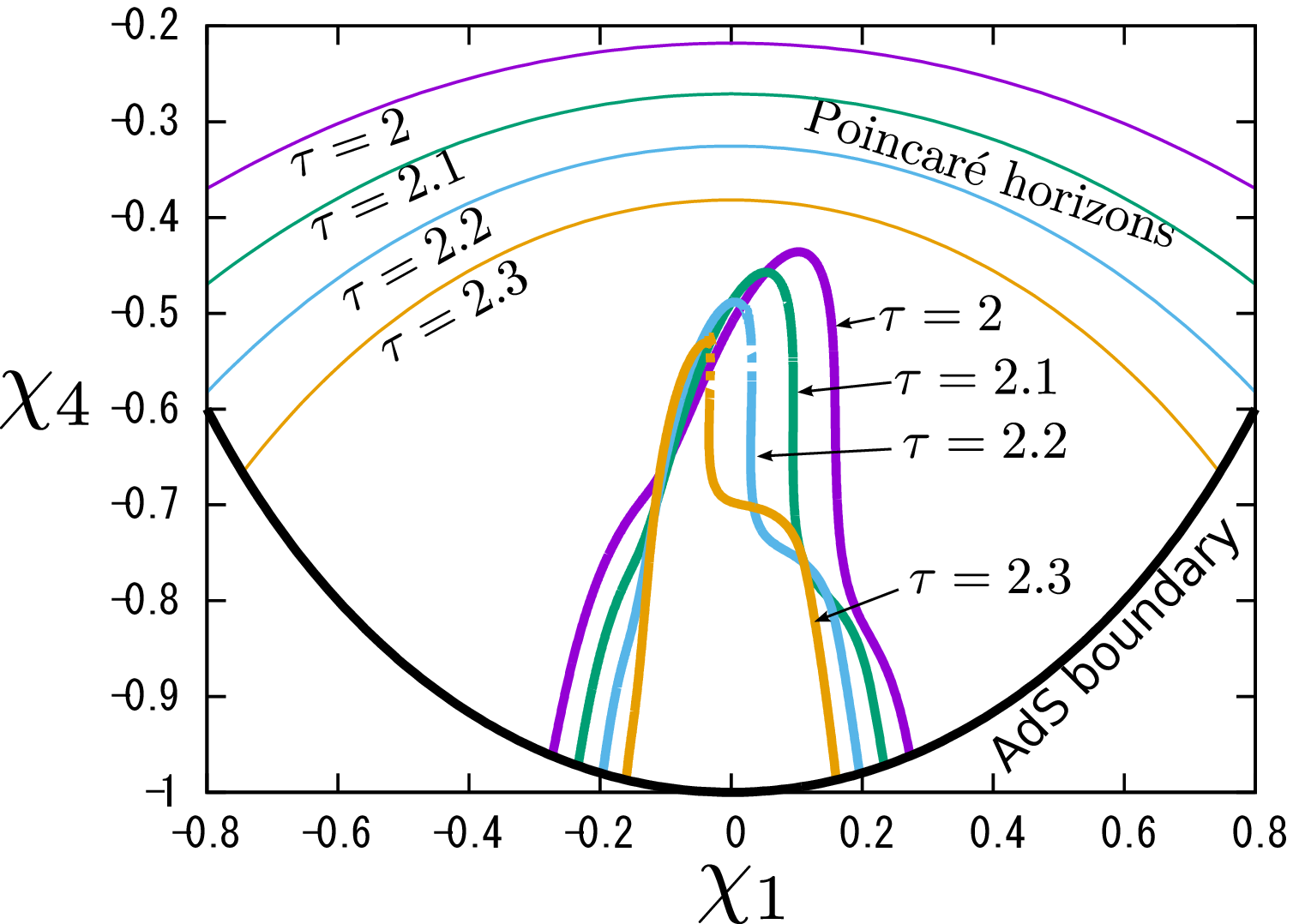}\label{LLearly}
}
\subfigure[$2.4\leq \tau \leq 2.7$]
{\includegraphics[scale=0.45]{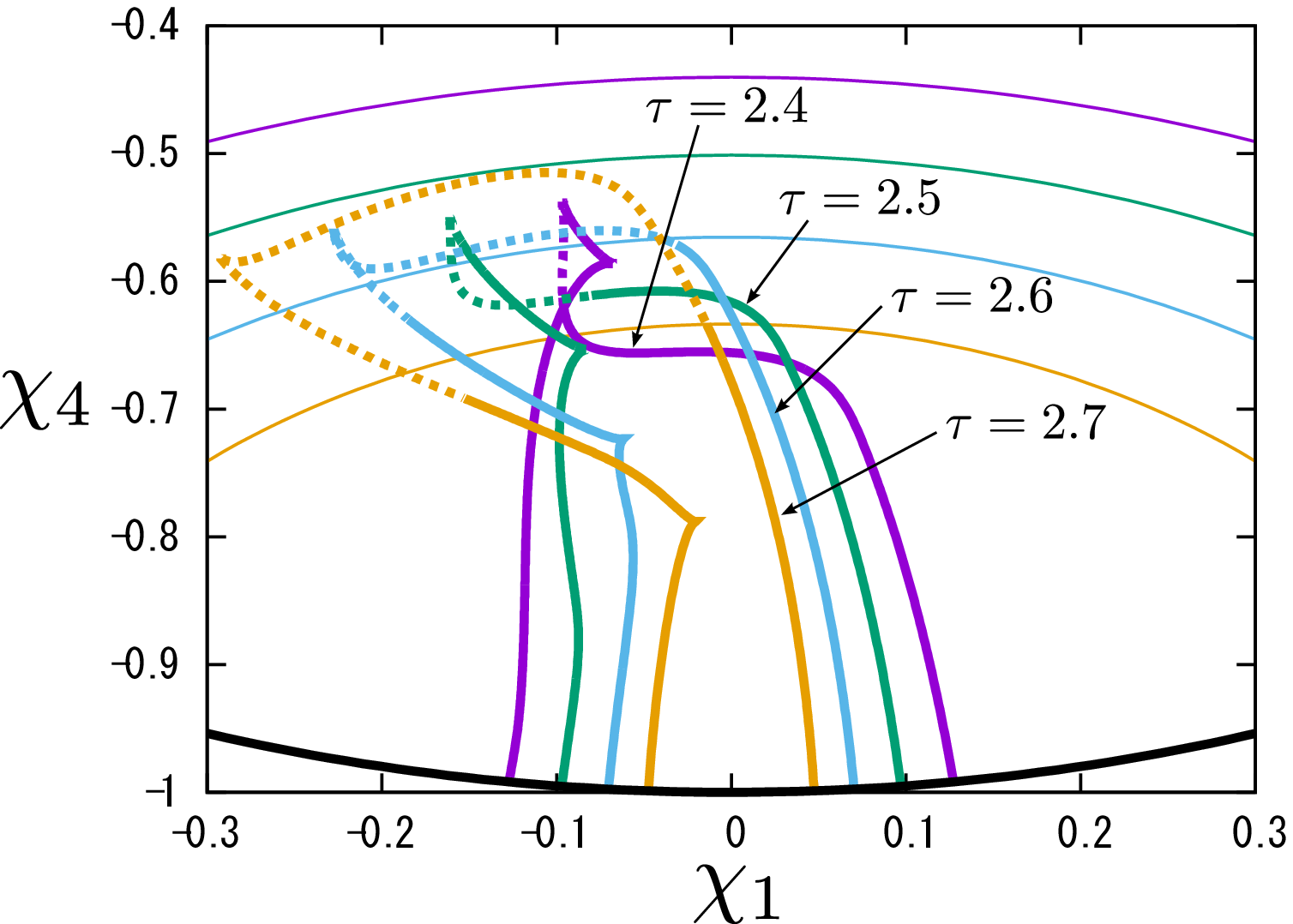} \label{LLlate}
}
\caption{%
More snapshots of the string dynamics in the global patch for the longitudinal quench with $\epsilon=0.15$ and $\Delta t/L=2$ for $\ell/L=1$.
Figs.~(a) and (b) are for $2\leq \tau \leq 2.3$ and $2.4\leq \tau \leq 2.7$, respectively.
The thick and thin curves correspond to the string configurations and Poincar\'e horizons.
The part of the string depicted by the dashed curve is inside of the effective event horizons on the worldsheet.
\label{LLsnap}
}
\end{figure}

We look into this dynamical string in Fig.~\ref{LLsnap}.
In these plots, thick and thin curves correspond to the string configurations and Poincar\'e horizons, respectively, and the part of the string inside of the worldsheet effective event horizons is 
depicted by the dashed curves. Note that for variables in the global coordinates nothing irregular happens.
For $2\leq \tau \leq 2.3$, we see that the string is smooth and does not reach the Poincar\'e horizon.
At $ \tau \geq 2.4$, cusps and self-intersection appear on the string.
We then observe that the string crosses the Poincar\'e horizon around $\tau\simeq 2.6$.
Prior to this time, the pair of the effective horizons appears around $\tau \simeq 2.1$.

\begin{figure}
\begin{center}
\includegraphics[scale=0.55]{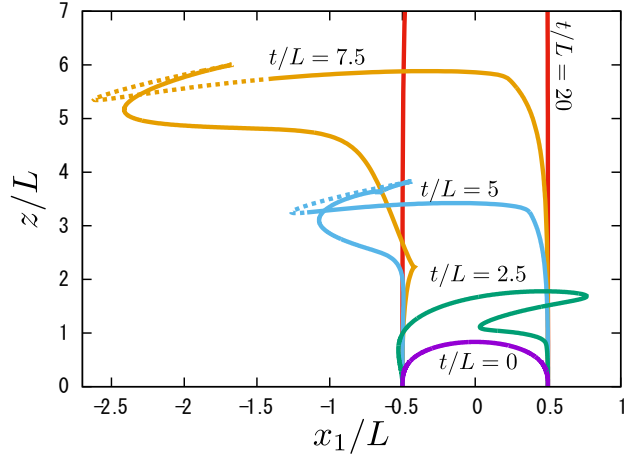}
\end{center}
\caption{%
Snapshots of the string dynamics in the Poincar\'e patch for the longitudinal quench with $\epsilon=0.15$ and $\Delta t/L=2$.
The effective horizons appear on the worldsheet within finite time.
The string configuration approaches two straight strings in the late time.
}
\label{LLpoin}
\end{figure}

This string motion in the global patch is seen as disconnection of the string in the Poincar\'e patch. 
The numerical solutions $\tau(u,v)$ and $\bm{\chi}(u,v)$ are mapped to the Poincar\'e patch's $t=T(u,v)$, $z=Z(u,v)$
and $\bm{x}=\bm{X}(u,v)$ through Eqs.~(\ref{GtoP_t}-\ref{GtoP_x}).\footnote{%
In the Poincar\'e patch, we use capital letters for functions specifying the string configuration following the notation in~\cite{Ishii:2015wua}.
}
In Fig.~\ref{LLpoin}, we show the corresponding string dynamics in the Poincar\'e patch.
The string becomes longer as $t$ increases, and in the late time its configuration eventually approaches two straight strings. We find that the effective event horizons appear around $t/L \simeq 2.54$.

While our time evolution is for purely classical strings, physically finite-$N_c$ effects can be important at the cusps and self-intersection points.
This implies that time evolution after their appearance may significantly alter due to such corrections, if they are formed before the event horizons.
However, as we will see in section \ref{sec:Trans},
in more general cases involving string dynamics in all $(4+1)$-dimensions,
no cusp formation nor self-intersection is found. Hence practically it is not likely that
the dynamical disconnection are prevented by the finite-$N_c$ effects in general quenches.

\subsection{Forces acting on the quarks}
\label{sec:force}

Forces act on the quark and antiquark endpoints as the response of the string.
In view of the gauge/gravity duality, the position of the string endpoint corresponds to the ``source'' of the field theory operator dual to the string, and that operator is the force. 
We compute the forces in the Poincar\'e patch, after the dynamical string solution is mapped from the global coordinates.
Let $x_i=X_i(t,z)$ denote the string coordinates in the target space.
The force on the quark is then given by \cite{Ishii:2015wua}
\begin{equation}
 F_i(t)=\frac{\sqrt{\lambda}}{4\pi} \gamma^{-1}(\delta_{ij}+\gamma^2 v_i v_j)\partial_z^3 X_j(t,z)|_{z=0, \bm{x}=\bm{x}_q}\ .
\label{force_formula}
\end{equation}
where the velocity $\bm{v}$ and Lorentz factor $\gamma$ have been defined in Eq.~(\ref{vgamma}).
The force on the antiquark can be simply given by changing $\bm{x}_q\to \bm{x}_{\bar{q}}$ with $\bm{v}=0$ and $\gamma=1$.

\begin{figure}[t]
\centering
\includegraphics[scale=0.45]{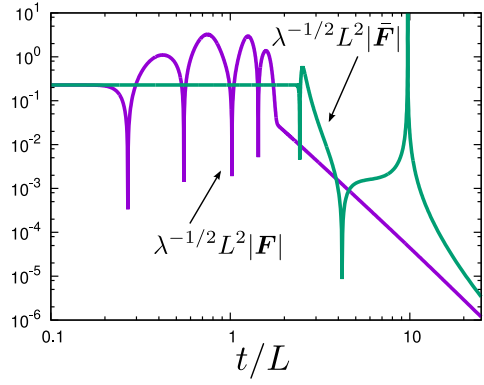}
\caption{The forces acting on the quark and antiquark 
for the longitudinal quench with $\epsilon=0.15$ and $\Delta t/L=2$.}
\label{FLL}
\end{figure}

In Fig.~\ref{FLL}, the absolute values of the forces for the longitudinal quench with $\epsilon=0.15$ and $\Delta t/L=2$ is shown in a log-log scale. In the plots, the forces on the quark and antiquark are denoted by $\bm{F}$ and $\bar{\bm{F}}$.
As $t \to \infty$, both $\bm{F}$ and $\bar{\bm{F}}$ approach zero.
This also indicates that the flux tube disconnects, and the final configuration (in $t \to \infty$) is two straight strings.
The divergence in $\bar{\bm{F}}$ at $t/L\sim 10$ is due to a cusp arriving at the boundary. (See~\cite{Ishii:2015wua}.)
Once the cusp passes, $\bar{\bm{F}}$ monotonically decays and its behavior seems to approach that of $\bm{F}$.

In the late time behavior of the forces, we observe power law decay. This is clearly seen in $\bm{F}$.
 In the case of $\epsilon=0.15$ and $\Delta t/L=2$, a fit of the power law in $\bm{F}$ is $t^{-n}$ with $n\sim3.8$.
It appears that $\bar{\bm{F}}$ also realizes power law decay, but because of its different evolution from $\bm{F}$ such as the travel of the cusp, the approach of $\bar{\bm{F}}$ to a power law seems quite delayed.

\begin{figure}[t]
\centering
\includegraphics[scale=0.45]{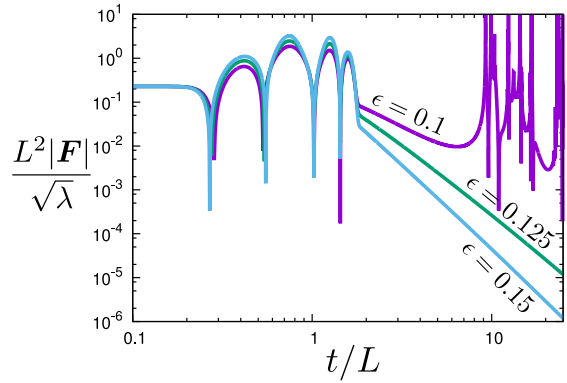}
\caption{Comparison of the forces acting on the quark for $\epsilon=$0.1, 0.125 and 0.15 when $\Delta t/L=2$.}
\label{powerlaw_dt2}
\end{figure}

The power depends on the quench parameters. In Fig.~\ref{powerlaw_dt2}, we compare $\bm{F}$ for $\epsilon=$0.1, 0.125, 0.15 when $\Delta t/L=2$. With the first value, the string extends its length a little but does not reach the Poincar\'e horizon, and the late time oscillations in the plot is due to bounces of cusps \cite{Ishii:2015wua}. For the latter two, the string reaches the Poincar\'e horizon. As seen in the $\epsilon=0.1$ plot, the power law can be observed even when the string does not reach the horizon, although it is disturbed by the travel of waves and cusps on the string. The power law exponent is read off $n \sim$ 2, 3.3, 3.8 for $\epsilon=$0.1, 0.125, 0.15, respectively. The decay becomes steeper as $\epsilon$ increases. The dependence on the quench parameters implies that the slope of the power law is not simply fixed by the AdS background.

What is the origin of the power law tail?
One of the most plausible explanations for the power law tail can be given
by the redshift of the effective horizon.\footnote{%
We first focus on the ``tail'' in the case that the string reaches the Poincar\'e horizon. The situation of the $\epsilon=0.1$ case is commented later.
}
Before going to detailed arguments, we emphasize that the curved initial configuration of the string is also important.
Let us define ``left moving'' as ``from the quark to antiquark'' ($\bm{x}_q \to \bm{x}_{\bar{q}}$) and ``right moving'' as ``from the antiquark to quark'' ($\bm{x}_{\bar{q}} \to \bm{x}_q$).\footnote{In Fig.~\ref{LLpoin}, we plotted $\bm{x}_{q}$ and $\bm{x}_{\bar{q}}$ endpoints to the right and left, respectively.}
While a left moving wave is mainly generated on the worldsheet by the quench, 
right moving waves are also generated secondarily from the backscattering by the curvature of the hanging-shape string and/or the reflection at the other boundary.
Such right moving waves then can propagate back to the quark endpoint.
Later in this section, we will discuss that starting from a straight string and adding boundary quenches do not result in producing a power law.

Suppose that the left and right moving waves are present on the string and effective event horizons are formed on the string. We then focus on the force on the quark endpoint.
Let us consider a right moving wave which has a non-trivial profile at the effective horizon.
A wave from near the horizon needs very long time to reach the AdS boundary 
and would cause the tail in the force.
For an asymptotically AdS black hole with finite surface gravity $\kappa$, 
perturbations decay with $\sim e^{-\kappa t}$.\footnote{%
For asymptotically AdS black holes with finite temperature,
there is no power law tail in perturbations
unlike asymptotically flat black holes~\cite{Horowitz:1999jd}.
}
In the present case, however, the effective horizon approaches the Poincar\'e horizon that can be regarded
as an extremal horizon ($\kappa=0$).
For an extremal case, the exponential decay is absent and replaced by a power law decay~\cite{Aretakis:2011ha,Aretakis:2011hc,Lucietti:2012xr,Murata:2013daa}.
The worldsheet effective horizons on the string in pure AdS might inherit such a property.

The presence of the power law decay of a perturbation can be demonstrated by a linear theory as follows.
(See also~\cite{Lucietti:2012xr} for the explicit calculation of the propagation of a scalar field in AdS$_2$.)
Let us consider a perturbation around a trivial solution of a straight string: $\bm{X}(t,z)=0$. 
The equation for the linearized perturbation is simply given by
\begin{equation}
\left(\partial_t^2-\partial_z^2+\frac{2}{z}\partial_z\right) \delta X_1(t,z)=0\ ,
\label{lineq}
\end{equation}
where we consider the perturbation only along the $x_1$-direction.
Its general solution is given by
\begin{equation}
 \delta X_1(t,z)=f_L(t-z)-f_R(t+z)+z\{\dot{f}_L(t-z)+\dot{f}_R(t+z)\}\ ,
\label{gensol}
\end{equation}
where
$f_L$ and $f_R$ are arbitrary functions which represent left and right moving waves.
If we put a compactly supported initial data as usually done in perturbation theory analyses,
we can never observe a power law tail. However, as seen in Fig.~\ref{LLpoin},
the string is bent even at the effective horizon in our numerical calculations.
For this reason, as a toy model that has non-trivial profile near the horizon,
we consider initial data, 
$\delta X_1|_{t=0}\sim 1/z^n$ $(z\to\infty)$ and $\delta \dot{X}_1|_{t=0}=0$, where $n$ is a non-negative constant. 
We also impose Dirichlet boundary condition at $z=0$.
Then, the arbitrary functions are determined as $f_R(x)=f_L(x)\sim 1/x^n$ $(x\to\infty)$.
Substituting them into Eq.~(\ref{gensol}) and
taking the limit of $t\to \infty$ for a fixed $z$, we obtain $\delta X_1\sim z^3/t^{n+3}$. 
Therefore, the force on the quark could obey a power law as $F\propto \partial_z^3 \delta X_1|_{z=0} \sim 1/t^{n+3}$.\footnote{%
Note that, for a horizon with finite surface gravity $\kappa$, 
the perturbation is always suppressed as $\sim e^{-\kappa t}$ even if we assume the power law initial data.}
Of course, in our actual numerical calculations, the profile of the string after the quench is not so simple as $\delta X_1|_{t=0}\sim 1/z^n$ but depends on details of how the string is deformed by the quenches. That would cause $\epsilon$ and $\Delta t$ dependence of the power law exponent. In particular, in our nonlinear numerical results $F \propto 1/t^n$ with $n<3$ is possible, which cannot be explained by the assumption in this paragraph's linear theory arguments.

As seen in Fig.~\ref{powerlaw_dt2}, the power law decay is temporarily present
even for $\epsilon=0.1$ and $\Delta t/L=2$ with which the string does not plunge into the Poincar\'e horizon.
This behavior can be observed roughly until the string stops expanding.
It would be fair to speculate that redshift effects on the stretching string might be generally responsible for causing the power law even if worldsheet event horizons are not formed.

The power law tail would be regarded as a relic of the dynamical ``phase transition'' between the hanging string and two straight strings. 
That is, the initial curved configuration of the string producing ``right-moving'' waves might be important.
Actually, we can never observe the power law tail if we perturb the endpoint of the straight string (single quark). 
Let us consider a linear perturbation again.
Imposing the boundary condition $\delta X_1|_{z=0}=x_q(t)$ and the initial condition $\delta X_1|_{t=0}=\delta \dot{X}_1|_{t=0}=0$,
we obtain a solution as $\delta X_1=x_q(t-z)+z\dot{x}_q(t-z)$.
This solution is purely ingoing wave and there is no backscattering. 
After the quench, therefore, the force on the quark becomes exactly zero and cannot have a power law tail. 
Although here we considered the linear perturbation for simplicity, 
the same consequence can be shown even for the full nonlinear solution obtained in~\cite{Mikhailov:2003er}, which applies to the Poincar\'e straight string.
We also computed nonlinear evolution of a straight string with boundary perturbations by using our numerical methods and checked that the force dropped to zero without a tail.

\subsection{Condition for the disconnection}

Stepping forward from the above example, we investigate dependence on the quench parameters. On one hand, we compute the energy of the dynamical string and evaluate the energy increase by quenches. On the other hand, we directly search critical parameters for the string to cross the Poincar\'e horizon. By comparing the results, we argue the condition for the string disconnection.

Let us derive the formula for evaluating the energy of the dynamical string.
In the global patch, the worldsheet action is given by (\ref{action1}).
Using Eq.~(\ref{PtoG}), we can rewrite the action in terms of the Poincar\'e coordinates as\footnote{%
While the energy can be equally calculated in the global coordinates, because of the simplicity of regularization we work in the Poincar\'e coordinates.
}
\begin{equation}
\begin{split}
S&=\frac{\sqrt{\lambda}}{2\pi}\int dudv\, \frac{1}{Z^2}(-T_{,u}T_{,v}+Z_{,u}Z_{,v}+\bm{X}_{,u}\cdot\bm{X}_{,v}) \\
&=\frac{\sqrt{\lambda}}{4\pi}\int d\sigma^0 d\sigma^1 \, \frac{1}{Z^2}\eta^{ab}(T_{,a}T_{,b}-Z_{,a}Z_{,b}-\bm{X}_{,a}\cdot\bm{X}_{,b})
\ ,
\end{split}
\end{equation}
where we introduced $\sigma^0=u+v$, $\sigma^1=u-v$, and $\eta^{ab}=\textrm{diag}(-1,1)$.
The domains of the coordinates are $-\infty<\sigma^0 <\infty$ and $0\leq \sigma^1 \leq \beta_0$.
This action is invariant under the time translation: $T\to T+\textrm{const}$.
Its conserved current is given by
\begin{equation}
 P^a=\frac{\sqrt{\lambda}}{2\pi} \frac{1}{Z^2}\eta^{ab}T_{,b}\ .
\end{equation}
To compute the string total energy, we integrate $P^0$ along $\sigma_1$ for fixed $\sigma_0$,\footnote{Since we consider only one-sided quenches, the fact that there is no energy inflow from the other endpoint makes the situation easy. We can simply use the constant-$\sigma_0$ slices for integration. If the other endpoint is also quenched, in order to define a total energy meaningful for the boundary field theory, it would be necessary to use spacelike slices where the two endpoints have the same boundary time.}
\begin{equation}
E_\textrm{bare} = -\int d\sigma^1 P^0 
= \frac{\sqrt{\lambda}}{2\pi}\int_\epsilon^{\beta_0-\epsilon'} d\sigma^1 \, \frac{\partial_0 T}{Z^2} \ .
\end{equation}
This quantity diverges in the limits of $\epsilon\to 0$ and $\epsilon'\to 0$ due to contributions at the AdS boundary.
The divergence is interpreted as coming from the energy of the infinitely heavy quarks $\sim m_q \gamma$ and $m_{\bar{q}}\bar{\gamma}$.
The energy is regularized by adding counter terms as
\begin{equation}
\begin{split}
E_\textrm{reg} &= E_\textrm{bare} -
\frac{\sqrt{\lambda}}{2\pi}\left[
\left.\frac{\gamma(T)}{Z}\right|_{\sigma^1=\epsilon}+\left.\frac{\bar{\gamma}(T)}{Z}\right|_{\sigma^1=\beta_0-\epsilon'}\right]\ .
\end{split}
\end{equation}
Substituting the static hanging string solution~(\ref{Pstatic}) into the above expression, we obtain the static energy as
$E_\mathrm{stat}L/\sqrt{\lambda}=-4\pi^2/\Gamma(1/4)^{4} \simeq -0.2285$,
consistent with the result in~\cite{Rey:1998ik,Maldacena:1998im}.
This value corresponds to the energy difference between the static hanging string and straight strings, where the regularized energy of the latter is zero.

In actual numerical calculations,
we include the boundary terms in the integrand as
\begin{equation}
 E_\textrm{reg} = \frac{\sqrt{\lambda}}{2\pi}\int^{\beta_0}_0 d\sigma^1\,\left[\frac{\partial_0 T}{Z^2}
-\partial_1\left(\frac{s(\sigma^1)}{Z}\right)\right]\ ,
\label{Ereg_integration_formula}
\end{equation}
where $s(\sigma^1)$ is any smooth function satisfying
$s=\gamma(T)|_{\sigma^1=0}+\mathcal{O}((\sigma^1)^2)$ ($\sigma^1\to 0$)
and $s=-\bar{\gamma}(T)|_{\sigma^1=\beta_0}+\mathcal{O}((\beta_0-\sigma^1)^2)$ ($\sigma^1\to \beta_0$).
This form is convenient because the integrand is finite at $\sigma^1=0$ and $\beta_0$.
The difference between the time dependent value of \eqref{Ereg_integration_formula} and the static value gives the amount of energy injected to the string by the quench until that moment. Evaluating the energy increase in $t \ge \Delta t$ gives the total addition of the energy.

\begin{figure}
\begin{center}
\includegraphics[scale=0.45]{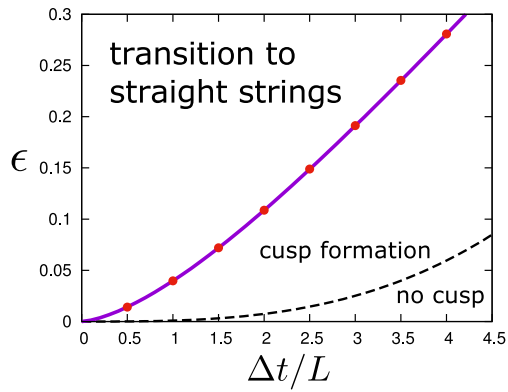}
\end{center}
\caption{Results of the evaluation of the string configuration change in $(\epsilon,\Delta t/L)$-parameters. The dots are the critical parameters obtained from the direct search. The real line corresponds to the energy injection equal to the energy difference of the two static string configurations. For comparison, an estimation of the threshold for cusp formation $\epsilon=9.3\times10^{-4}(\Delta t/L)^3$ obtained in \cite{Ishii:2015wua} is also plotted with the dashed line.}
\label{phase_epsdt}
\end{figure}

By changing $\epsilon$ and $\Delta t$, we compute the energies injected to the string. In particular, we are interested in quenches where the total energy injection is equal to the energy difference between the straight and hanging strings. We obtain the values of $(\epsilon,\Delta t)$ for such critical injections. The results will be soon discussed in Fig.~\ref{phase_epsdt}.

An alternative way to evaluate the amount of the energy injection is to compute the work done on the quark endpoints. The total work is given by\footnote{In~\cite{Ishii:2015wua}, it was argued that an extra term proportional to $\partial_t (\gamma \bm{v})$ may appear in the regularized force \eqref{force_formula}. However, that term becomes $\bm{v}\cdot \partial_t (\gamma \bm{v}) = \partial_t \gamma$ in the integrand, and since $\gamma=0$ at the boundaries of the integration ($t=0, \, \Delta t$), there is no contribution in \eqref{work_formula}.}
\begin{equation}
W = -\int_{C} d \bm{x} \cdot \bm{F} = -\int_0^{\Delta t} dt \, \bm{v} \cdot \bm{F}\ ,
\label{work_formula}
\end{equation}
where in the last expression we used the fact that our quench patterns are compact in time. The integrand $\bm{v}\cdot\bm{F}$ has the form of the product of time derivative of the ``source'' and the ``response'' in the sense of holography and reflects non-conservation of the energy in the presence of a time dependent source. We checked that the total work computed by evaluating \eqref{work_formula} agrees with the energy change evaluated from \eqref{Ereg_integration_formula}.

We also directly search for critical quench parameters for the string to cross the Poincar\'e horizon. For a fixed $\Delta t$, we start from a large amplitude and measure $\tau$ when the string crosses the horizon. As the amplitude is decreased, the crossing time approaches $\tau=\pi$, but computations also become difficult around $\chi_4=-1$. While we can compute for crossing times very close to $\tau = \pi$, once numerical errors become significant, we use an extrapolation to estimate the critical amplitude with which the crossing time would be $\tau=\pi$. We repeat this by changing $\Delta t$ and obtain critical $(\epsilon,\Delta t/L)$ 

In Fig.~\ref{phase_epsdt}, we plot the results of the searches. The result of the energy injection equal to the difference between the two string configurations (i.e.~when Eq.~\eqref{Ereg_integration_formula} gives $E_\textrm{reg}=0$) is plotted with a line.
That from the direct search for the horizon crossing is plotted with dots.
Remarkably, these two results agree. This implies that if the amount of the energy larger than the static energy is injected, the hanging string dynamically changes its topology to two straight strings.
By this process, the two endpoints are causally disconnected. This phenomenon is seen in the dual field theory as a breaking of the flux tube when the energy bigger than the ``binding'' energy is supplied.

We can show that the string cannot plunge into the Poincar\'e horizon for $E_\textrm{reg}<0$ as follows.
Let us consider dynamical solutions intersecting with the Poincar\'e horizon.
In terms of Poincar\'e target space coordinates, the regularized energy for such a string is given by
\begin{equation}
 E_\textrm{reg}=\frac{\sqrt{\lambda}}{2\pi}\int^\infty_0 dz \left[\frac{1+\bm{X}'{}^2}{z^2 \sqrt{(1-\dot{\bm{X}}^2)(1+\bm{X}'{}^2)-(\dot{\bm{X}}\cdot\bm{X}')^2}}-\frac{1}{z^2}\right]\ ,
\end{equation}
where we took the upper bound of the integration as $z=\infty$ since the string intersects with the horizon, and $\dot{\bm{X}} \equiv \partial_t \bm{X}$ and $\bm{X}' \equiv \partial_z \bm{X}$.
It is straightforward to check that the straight static string $\bm{X}(t,z)=0$ minimizes the above energy as $E_\textrm{reg}=0$. 
This implies that, even if time-dependence is taken into account, string configurations with $E_\textrm{reg}<0$ cannot intersect with the horizon.

Does a string with $E_\textrm{reg}>0$ always intersect with the Poincar\'e horizon?
This is a difficult question since to obtain an answer it is necessary to consider all time dependent solutions with positive energies.
At least in our numerical calculations, the final fate of the string dynamics for $E_\textrm{reg}>0$ was always straight strings.
The straight strings may be regarded as an attractor for positive energy solutions. 
It would be interesting to see if it is possible to find fine-tuned initial data and boundary conditions by which a
string never plunges into the horizon.

\section{Transverse quenches}
\label{sec:Trans}

\begin{figure}
\centering
\subfigure[Global patch]
{\includegraphics[scale=0.5]{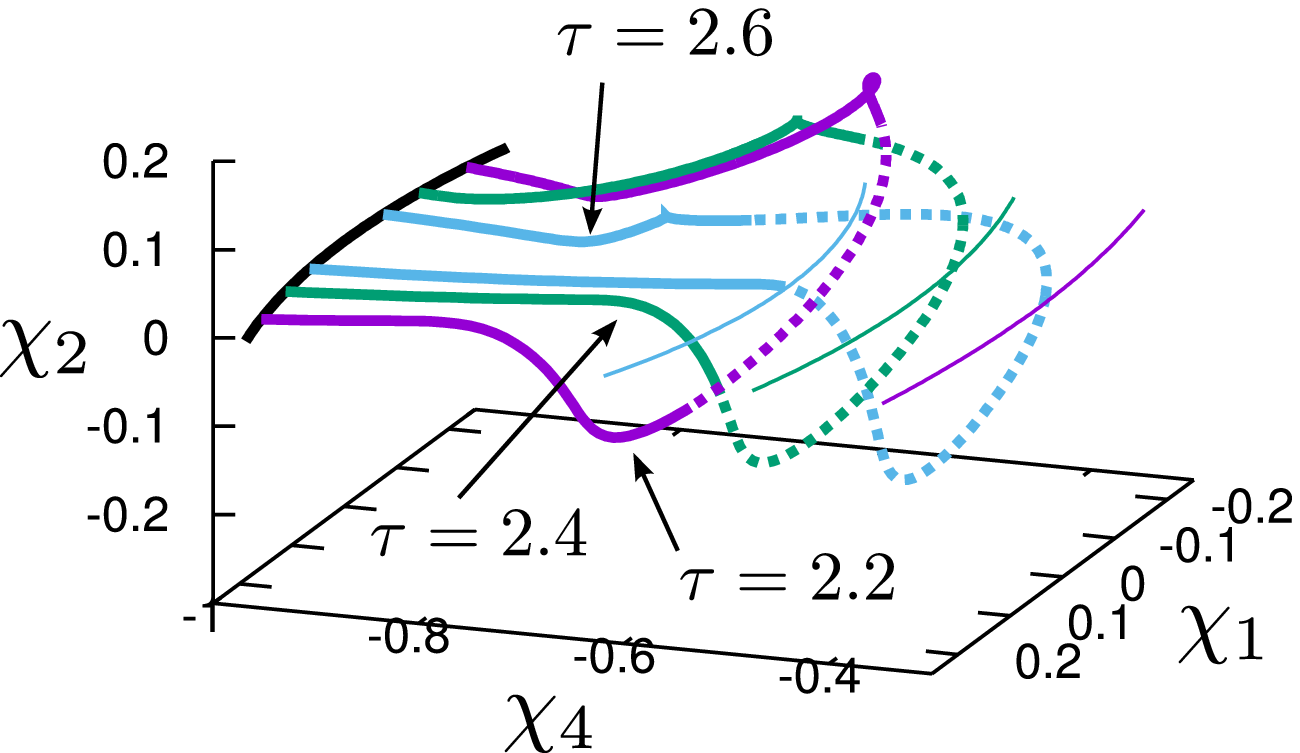}\label{TLglob}
}
\subfigure[Poincar\'e patch]
{\includegraphics[scale=0.5]{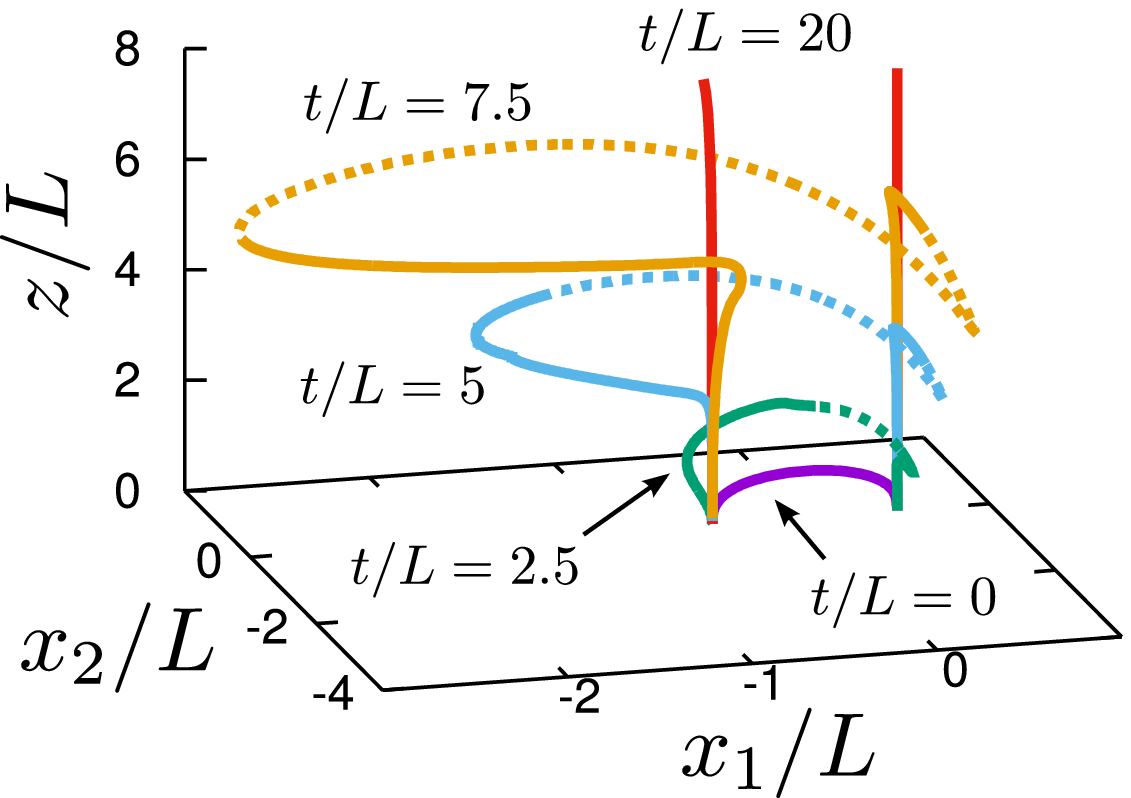} \label{TLpoin}
}
\caption{%
Snapshots of the string dynamics for the transverse linear quench with $\epsilon=0.2$ and $\Delta t/L=2$.
Figs.~(a) and (b) are for the global and Poincar\'e patches, respectively, and Fig.~(a) is for $\ell/L=1$.
In Fig.~(a), Poincar\'e horizons are shown by thin curves only on the ($\chi_1,\chi_4$)-plane for visibility.
Dashed curves 
correspond to the region inside the effective event horizons.
\label{TLsnap}
}
\end{figure}

We also consider transverse quenches where string motions are in more spatial dimensions than the longitudinal quenches. 
In Fig.~\ref{TLsnap}, we show snapshots of the string dynamics 
for the transverse linear quench \eqref{quench_tranlin} with $\epsilon=0.2$ and $\Delta t/L=2$.
The string motion is in the ($3+1$)-dimensions spanned by ($t,\chi_1,\chi_2,\chi_4$).
Fig.~\ref{TLglob} visualizes that the string plunges into the Poincar\'e horizon accompanied by the effective horizon formation on the worldsheet.
The solution mapped to the Poincar\'e patch is shown in Fig.~\ref{TLpoin}.
The string becomes longer as $t$ increases and
in the late time approaches the configuration with two straight strings.
Although it is not so much clear in Fig.~\ref{TLsnap},
we found cusp formation by checking roots of the Jacobian $\tau_{,u}\bm{\chi}_{,v}-\tau_{,v}\bm{\chi}_{,u}$ \cite{Ishii:2015wua}.
On the other hand, we did not observe self-intersections of the string in this example although 
it is possible from the dimensional point of view for the transverse linear quench.
Self intersections of the string could be observed in general ($3+1$)-dimensional string dynamics.

\begin{figure}
\begin{center}
\subfigure[$(x_1,x_2,z)$-space]
{\includegraphics[scale=0.5]{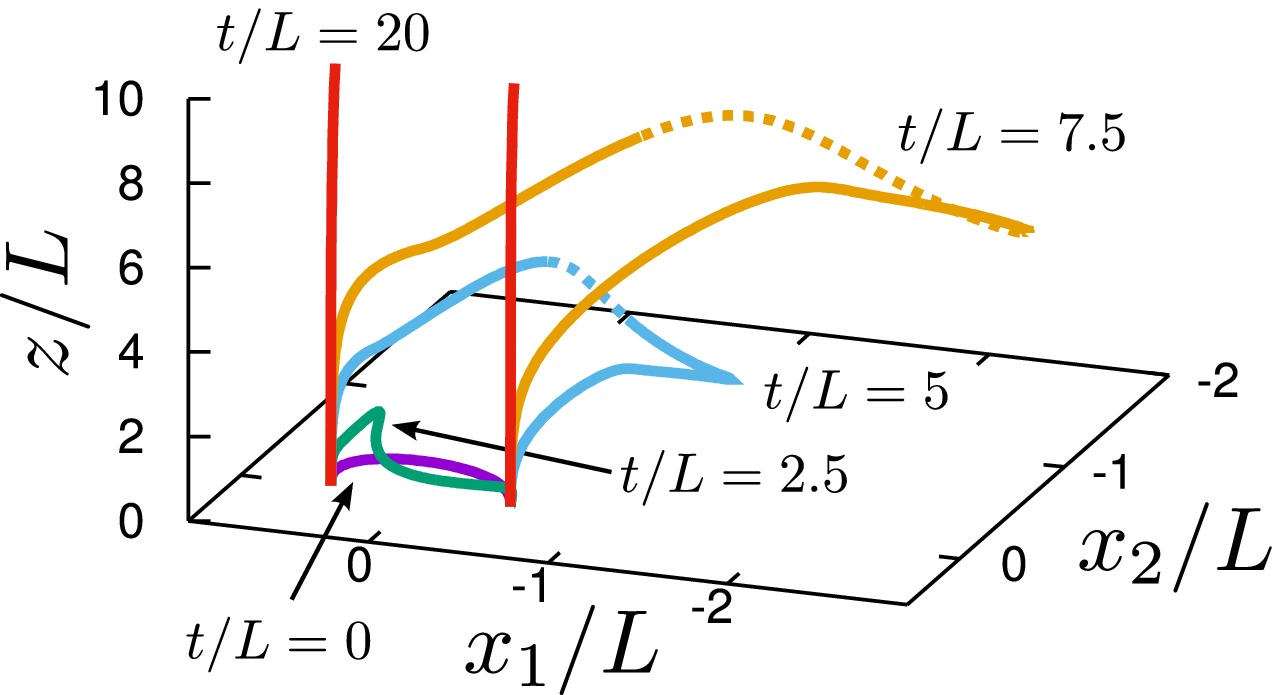} \label{TCx1x2z}
}
\subfigure[$(x_1,x_3,z)$-space]
{\includegraphics[scale=0.5]{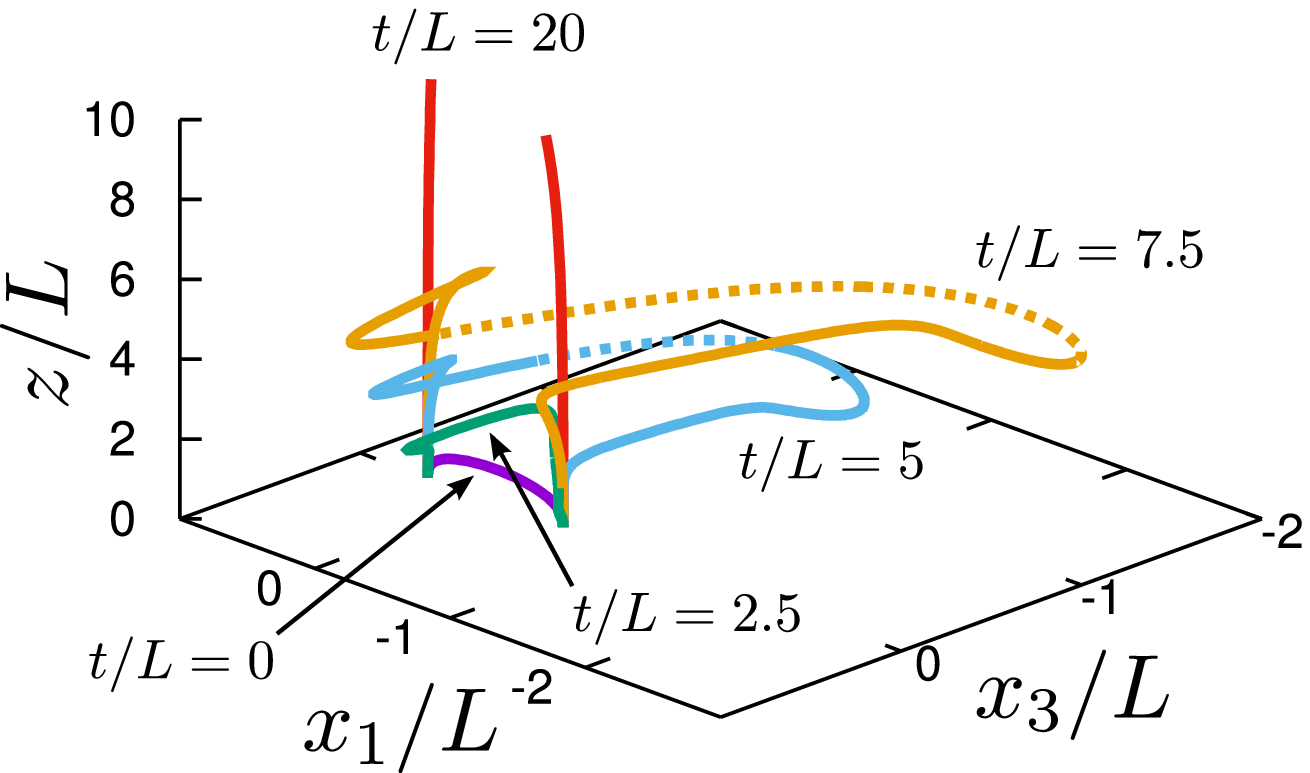} \label{TCx1x3z}
}
\subfigure[$(x_1,x_2,x_3)$-space]
{\includegraphics[scale=0.5]{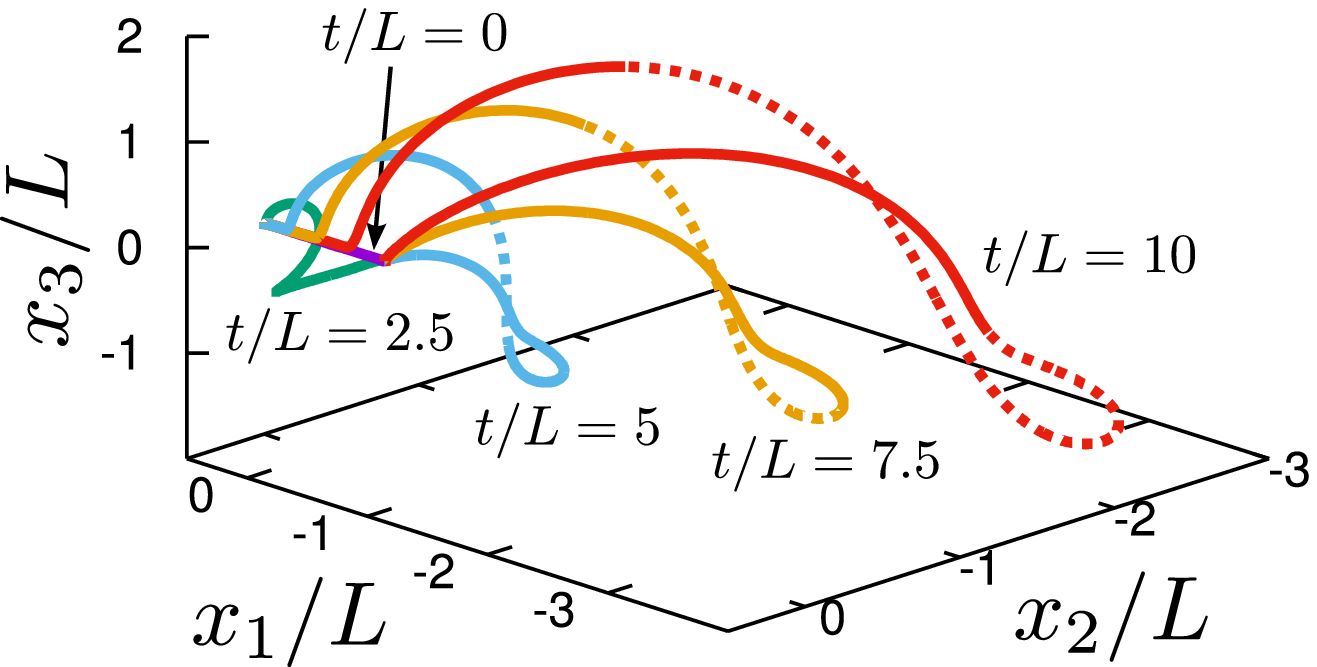} \label{TCx1x2x3}
}
\end{center}
\caption{%
Poincar\'e coordinate snapshots of the string dynamics for the transverse circular quench with $\epsilon=0.1$ and $\Delta t/L=2$.
}
\label{TCsnap}
\end{figure}

Transverse circular quenches induce string dynamics in the whole AdS$_5$ spacetime.
Once we project the profile of the string into subspaces, ($\tau,\chi_1,\chi_2,\chi_3$) and ($\tau,\chi_1,\chi_2,\chi_4$), 
we find dynamics qualitatively similar to transverse linear quenches. 
However, we find no cusp formation nor self interaction.
In Fig.~\ref{TCsnap}, we show snapshots in the Poincar\'e coordinates by projecting the $(1+4)$-dimensional dynamics to three dimensional spaces.
The effective horizons appear on the string, and the string approach to $z\to\infty$ in the late time.
This implies that the quark and antiquark are causally disconnected by the strong quench also in this case.

\begin{figure}
\centering
\subfigure[Transverse linear quench]{\includegraphics[scale=0.45]{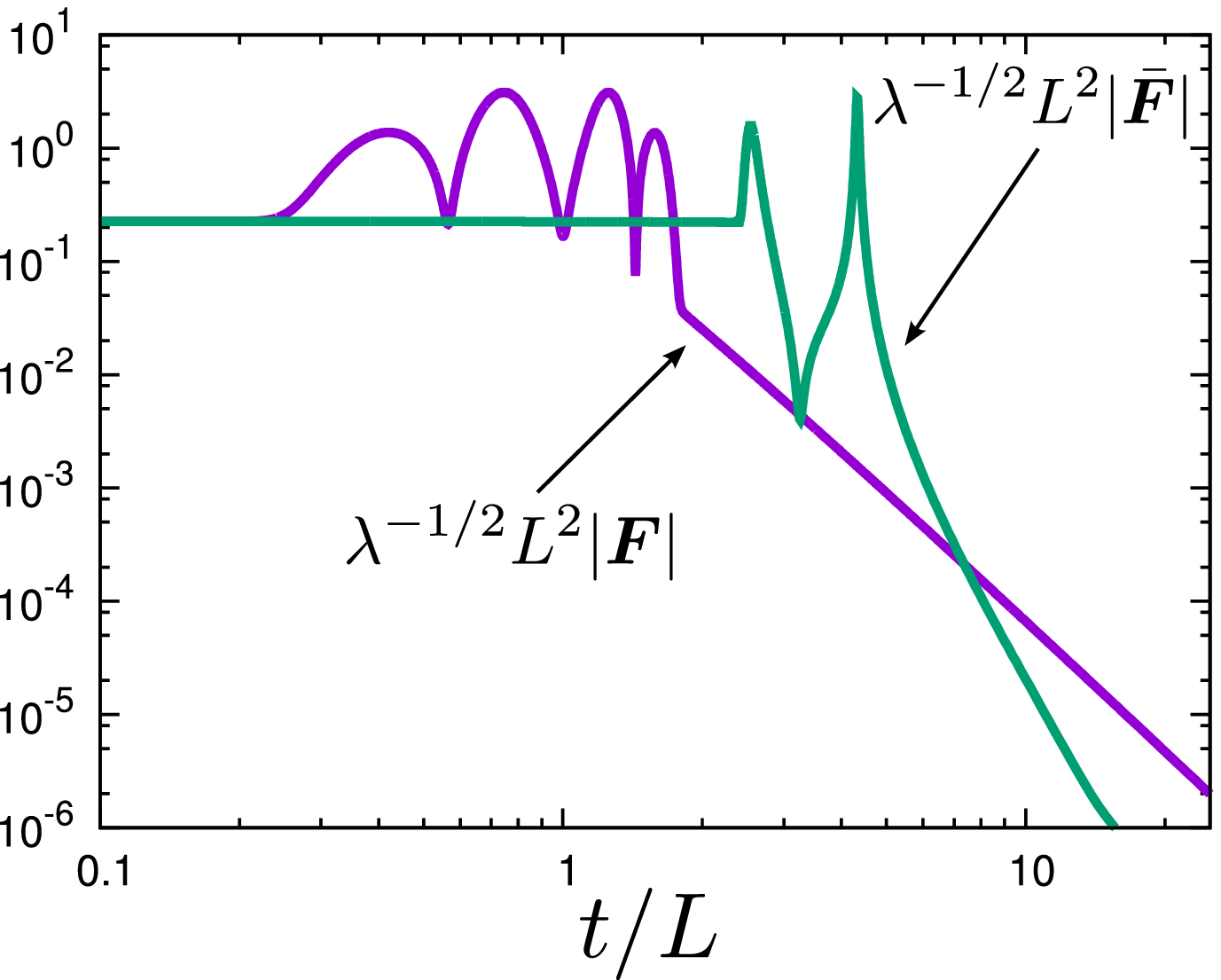}\label{TLforce}}
\hspace{2ex}
\subfigure[Transverse circular quench]{\includegraphics[scale=0.45]{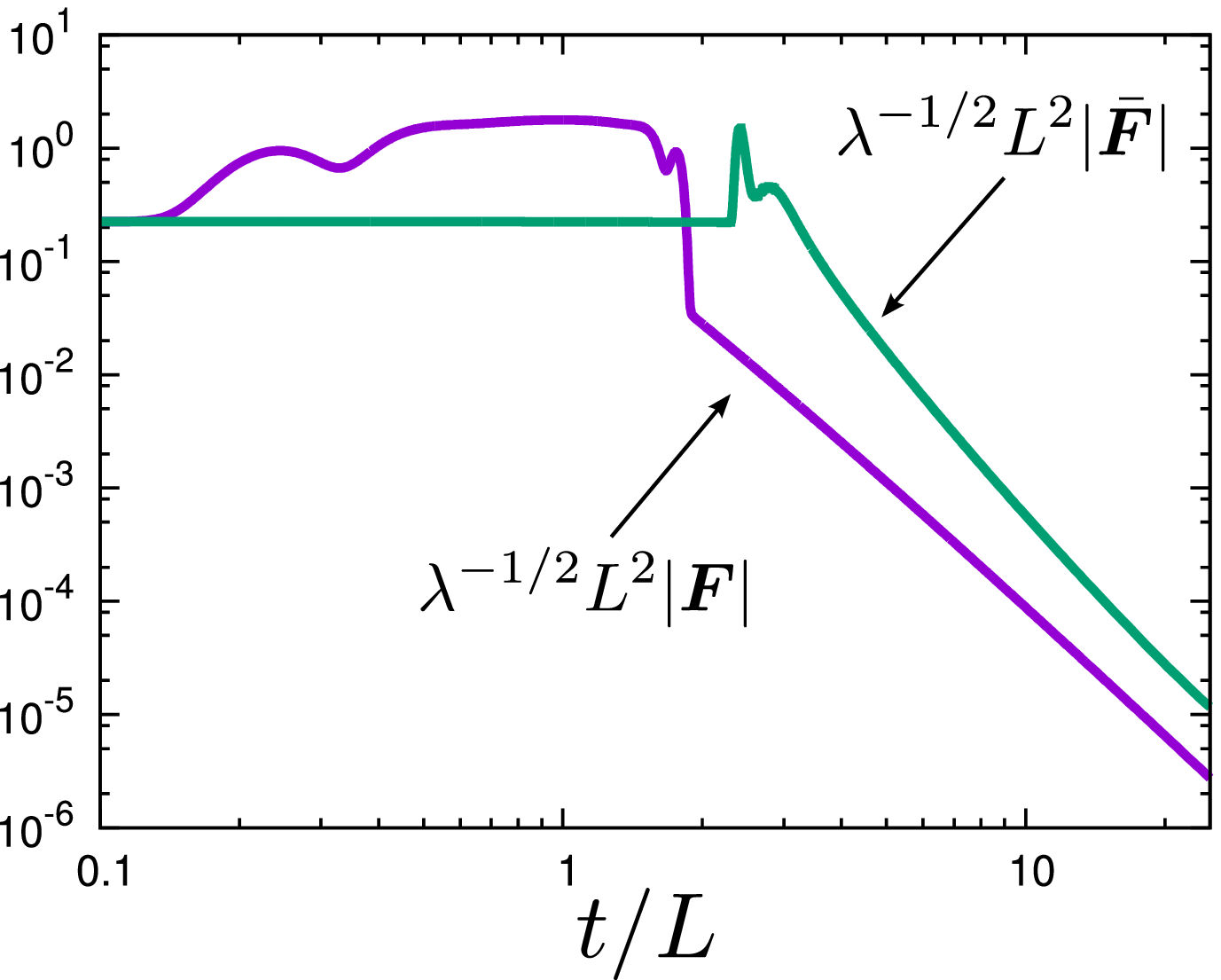} \label{TCforce}}
\caption{%
Forces acting on the quark and antiquark for the transverse quenches.
The parameters are $(\epsilon,\Delta t/L)=(0.15,\,2)$ and $(0.1,\,2)$ for the transverse linear and circular quenches, respectively. 
}
\label{Ftransforce}
\end{figure}

In Fig.~\ref{Ftransforce}, the forces acting on the quark and antiquark for the transverse quenches are shown. Their behaviors are similar to the case of longitudinal quenches: The forces show power law decay in the late time as the string expands. For the transverse linear quench with $(\epsilon,\Delta t/L)=(0.15,2)$, we find that the increase of the energy by the quench is almost the same as the case of the longitudinal quench with the same parameters. The power law in $\bm{F}$ is also found $t^{-3.8}$, consistent with the corresponding longitudinal quench. The transverse circular quench shown here has a little bit smaller energy injection and a different exponent $t^{-3.7}$. The force on the antiquark also seem to behave in the same power law as the quark side. In Fig.~\ref{TLforce}, however, the curve of $|\bar{\bm{F}}|$ does not look to approach that of $|\bm{F}|$, but this might be because of distortion by the cusps. In contrast, Fig.~\ref{TCforce} suggests that the behaviors of the forces will be similar in very late time.

In these examples of different quenches, we observe common transition to the power law tail region, and the string plunges into the Poincar\'e horizon once a sufficient amount of energy is supplied by the quench. Hence it looks that the the dynamical change of the string configurations is most likely governed by the strength of the perturbation and less sensitive to the details like quench patterns of dimensionality of string motions.

\section{Summary and discussion}
\label{sec:sum}

We studied strong perturbations of a fundamental string in AdS dual to the flux tube between a pair of external quark and antiquark in $\mathcal{N}=4$ super Yang-Mills theory. Non-linear perturbations were introduced to the string by shaking its endpoints. While our physical interest were in the Poincar\'e patch,
we adopted the global patch as the target space for solving the string dynamics numerically. 
We found that the string plunged into the Poincar\'e horizon if the perturbations were strong enough. 
In this process, effective event horizons were dynamically created on the worldsheet before the string reached the Poincar\'e horizon.
The condition for the string plunge was that the injected energy by a quench exceeded the potential energy of the $q$-$\bar{q}$ pair.
The forces acting on the quark and antiquark were found to approach zero at late time with power law decay, whose exponent depended on the quench parameters.
We argued that the presence of the power law tail might be associated with redshift effects on the string and the initial curved configuration of the string. Using the case that the string extended to the Poincar\'e horizon, we discussed the relationship between the power law and effects from the worldsheet effective horizons.

The slope of the power law depended on the quench parameters. As $\epsilon/\Delta t$ decreases, the exponent seems to be smaller, while it becomes difficult to observe a long-time power law tail because the string stop expanding and waves reflect between the boundaries. In the limit of $\epsilon/\Delta t \to 0$, no dynamics is induced on the string, and the exponent may also approach zero. On the other hand, for large $\epsilon/\Delta t$, the slope would be steeper. The velocity of the string endpoint by quenches reaches the speed of light at some finite $\epsilon/\Delta t$ \cite{Ishii:2015wua}. In such a situation, the bulk string element adjacent to the boundary is causally disconnected, and the initial shape of the bulk string would be irrelevant to the force. In that limit, the force may drop to zero suddenly after the compact quench is over.

One of the future directions of our work is to study string dynamics in AdS black hole background.\footnote{A related work in Vaidya-AdS is \cite{Ali-Akbari:2015ooa}.}
In finite temperatures, straight strings extending to the black hole represent deconfined quarks in gluon plasma \cite{Rey:1998bq,Brandhuber:1998bs}. It would be straightforward to see a nonlinarly perturbed hanging string change to straight strings extending to the black hole.
In general, it is known that perturbations around AdS black holes do not show power law decay tails \cite{Horowitz:1999jd}.
However, while the string is heading for the black hole, power law decay 
would appear, at least for a sufficiently low temperature, although it may be eventually taken over by an exponential ring down due to the presence of the black hole.
It would be interesting to see if a time domain of power law decay can be observed in high temperature black holes.
In application of the gauge/gravity duality to strongly coupled QCD, much attention has been also given to moving quarks in Yang-Mills plasma \cite{Herzog:2006gh,Gubser:2006bz,CasalderreySolana:2006rq,Gubser:2006nz,CasalderreySolana:2007qw,Liu:2006ug,Liu:2006he} and moving quark-antiquark pairs \cite{Liu:2006nn,Chernicoff:2006hi}. It would be interesting to apply our computational methods to study nonlinear fluctuations on quarks and mesons in strongly coupled plasmas from the viewpoint of the holographic string.

The interpretations of the disconnection of the string in the dual field theory 
is that the flux tube between the quark and antiquark vanish. While horizons are created on the string as if the string were broken into two, this actually is not analogous to the breaking of the flux tube in ``confined'' theories. In fact, $\mathcal{N}=4$ super Yang-Mills theory is conformal and not confining.
What we observed in our AdS setup was that the AdS string having two static solutions transitioned from the hanging to straight strings.
While this dynamical transition is calculated and observed in a simple setup of the pure AdS background in this work, this finding would have interesting implications in generalizations of this work: nonlinear perturbations of strings in confining geometries.
The string may become infinitely long in
confining geometries where no disconnected string configuration exists. A related study was done in \cite{Vegh:2015yua} for a global AdS string whose boundary conditions are different from ours, and a phase with endlessly extending string was found when perturbations were continuously added. It would be interesting to examine if a finite perturbation can make a string stretch forever as well as create event horizons on the string.

An exact solution of an expanding string with accelerating endpoints has been found in~\cite{Xiao:2008nr}.
This solution, having effective horizons on its worldsheet, has been regarded as the holographic dual of an Einstein-Podolsky-Rosen (EPR) pair~\cite{Jensen:2013ora} since the quark and antiquark pair should be color singlet (entangled)
but is causally disconnected (See also \cite{Maldacena:2013xja}).
In this paper, we found the dynamical horizon creation on the string worldsheet just by changing the string endpoints temporarily, and as a result the endpoints became causally disconnected. How the effective horizons shut their communication were schematically visualized in Fig.~\ref{effhor}. It would be interesting to consider entanglement and EPR pairing in this kind of dynamical horizon formation.

\acknowledgments
The authors thank Hans Bantilan, Paul Romatschke, and Kentaroh Yoshida for fruitful discussions and comments.
The work of T.I.~was supported by the Department of Energy, DOE award No.~DE-SC0008132.
The work of K.M. was supported by JSPS KAKENHI Grant Number 15K17658.

\appendix

\section{Discretization at the boundary}
\label{sec:disc}

In this appendix, we explain the discretized evolution at the boundary in the global coordinates. This is different from the Poincar\'e patch's flat boundary considered in \cite{Ishii:2015wua}. While the $\chi$-coordinates are useful for solving equations in the bulk, we find it convenient to go through the polar coordinates in deriving the discretized equations at the boundary. The coordinate change between them is given by
\begin{equation}
\chi_a = \omega_a \tan \frac{\theta}{2}\ ,
\label{chi_to_polar}
\end{equation}
where $0 \le \theta \le \pi/2$ is the AdS radial coordinate and $\omega_a$ are the spherical coordinates of $S^3$. In terms of the polar coordinates, the constraint equations at the boundary \eqref{bdryTevo_chi} become
\begin{align}
\tau_{,u} = \sqrt{\theta_{,u}{}^2 + |\bm{\omega}_{,u}|^2} \ , \quad
\tau_{,v} = \sqrt{\theta_{,v}{}^2 + |\bm{\omega}_{,v}|^2} \ .
\label{polar_boundary_taueq}
\end{align}
Let us discretize \eqref{polar_boundary_taueq}. For simplicity, we consider one of the string's boundaries at $u=v$; the other ($u=v+\beta_0$) can be handled similarly. Introducing notations $\phi_N = \phi(u+h,v+h), \, \phi_E = \phi(u,v+h), \, \phi_W = \phi(u+h,v)$ and $\phi_S = \phi(u,v)$ where $\phi$ represents the fields, we can discretize the sum of the two equations in \eqref{polar_boundary_taueq} as
\begin{multline}
\tau_N-\tau_S=\frac{1}{2}\left(
\sqrt{(\theta_N-\theta_E+\theta_W-\theta_S)^2+|\bm{\omega}_N-\bm{\omega}_E+\bm{\omega}_W-\bm{\omega}_S|^2} \right.\\
\left. +  \sqrt{(\theta_N-\theta_W+\theta_E-\theta_S)^2+|\bm{\omega}_N-\bm{\omega}_W+\bm{\omega}_E-\bm{\omega}_S|^2}\right).
\label{tau_disc_polar0}
\end{multline}
From the boundary condition for $\theta$, $\theta|_{u=v}=\pi/2$,
we have $\theta_N=\theta_S=\pi/2$.
By examining the boundary series expansions,
we find $\theta_{,uv}|_{u=v}=0$ and $\bm{\omega}_{,u}|_{u=v}=\bm{\omega}_{,v}|_{u=v}$.
Discretizing them, we obtain $\theta_E=\pi-\theta_W$ and $\bm{\omega}_E=\bm{\omega}_W$.
Therefore, \eqref{tau_disc_polar0} becomes
\begin{align}
\tau_N &= \tau_S + \sqrt{(\pi-2\,\theta_W)^2+|\bm{\omega}_N-\bm{\omega}_S|^2} \nonumber \\
&= \tau_S + \sqrt{(\pi-4 \tan^{-1}\!|\bm{\chi}_W|)^2+|\bm{\chi}_N-\bm{\chi}_S|^2}\ ,
\label{tau_disc_polar}
\end{align}
where we used $|\bm{\omega}_N-\bm{\omega}_S| = |\bm{\chi}_N-\bm{\chi}_S|$ on the AdS boundary and $\theta_W = 2 \tan^{-1}\!|\bm{\chi}_W|$ obtained from \eqref{chi_to_polar}.
By solving \eqref{tau_disc_polar} with \eqref{bdrycond_chi} giving $\bm{\chi}_N = \bm{\chi}(\tau_N)$, we can compute the time evolution, $\tau_N$ and $\bm{\chi}_N$.

\section{Error analysis}
\label{sec:error}

In this appendix, we estimate numerical errors by checking the constraint violation.
The constraint equations have been obtained in Eq.~(\ref{CON}).
To remove the coefficients diverging at the boundary, we define rescaled constraints as
\begin{equation}
 \tilde{C}_1=-(1+|\bm{\chi}|^2)^2 \, \tau_{,u}^2 + 4 |\bm{\chi}_{,u}|^2 \ ,\quad
 \tilde{C}_2=-(1+|\bm{\chi}|^2)^2 \, \tau_{,v}^2 + 4 |\bm{\chi}_{,v}|^2 \ .
\end{equation}
Mathematically, they should be zero in the whole computational region if we impose the constraints at the string's boundaries and initial surface.
Hence, they can be used for checking our numerical accuracy after discretization.

\begin{figure}
\centering
\subfigure[Longitudinal]
{\includegraphics[scale=0.3]{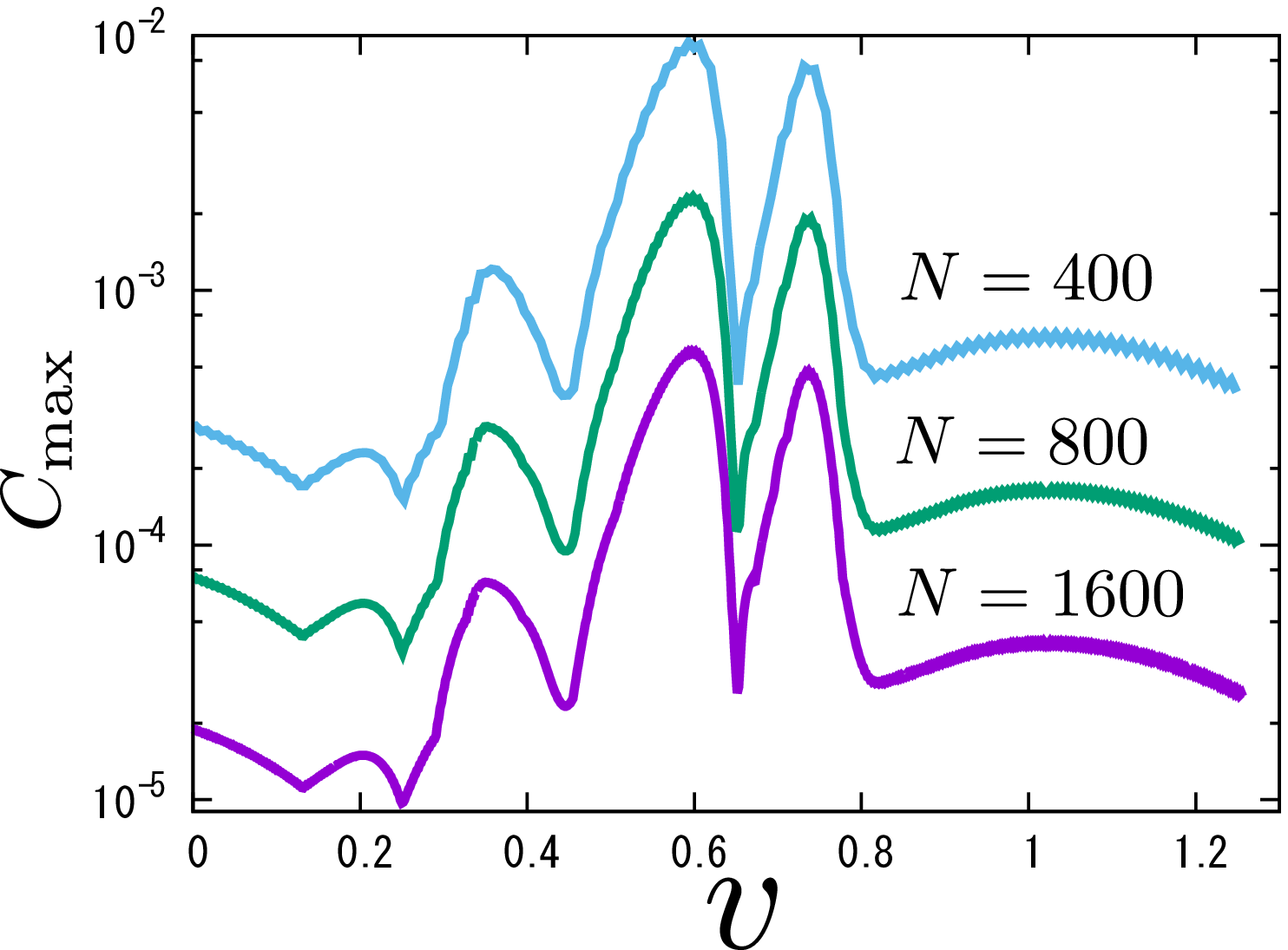}\label{CLL}
}
\subfigure[Transverse linear]
{\includegraphics[scale=0.3]{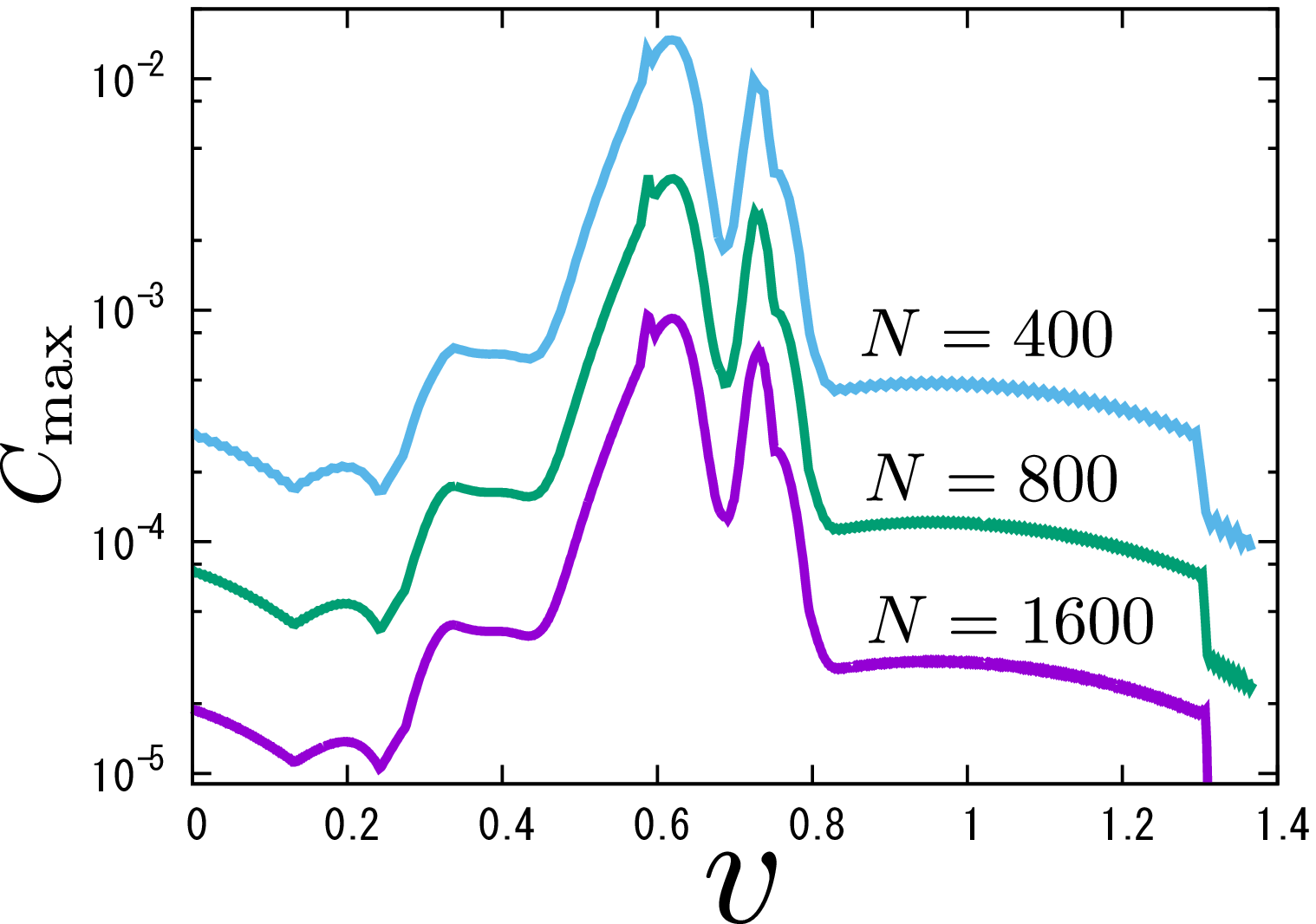}\label{CTL}
}
\subfigure[Transverse circular]
{\includegraphics[scale=0.3]{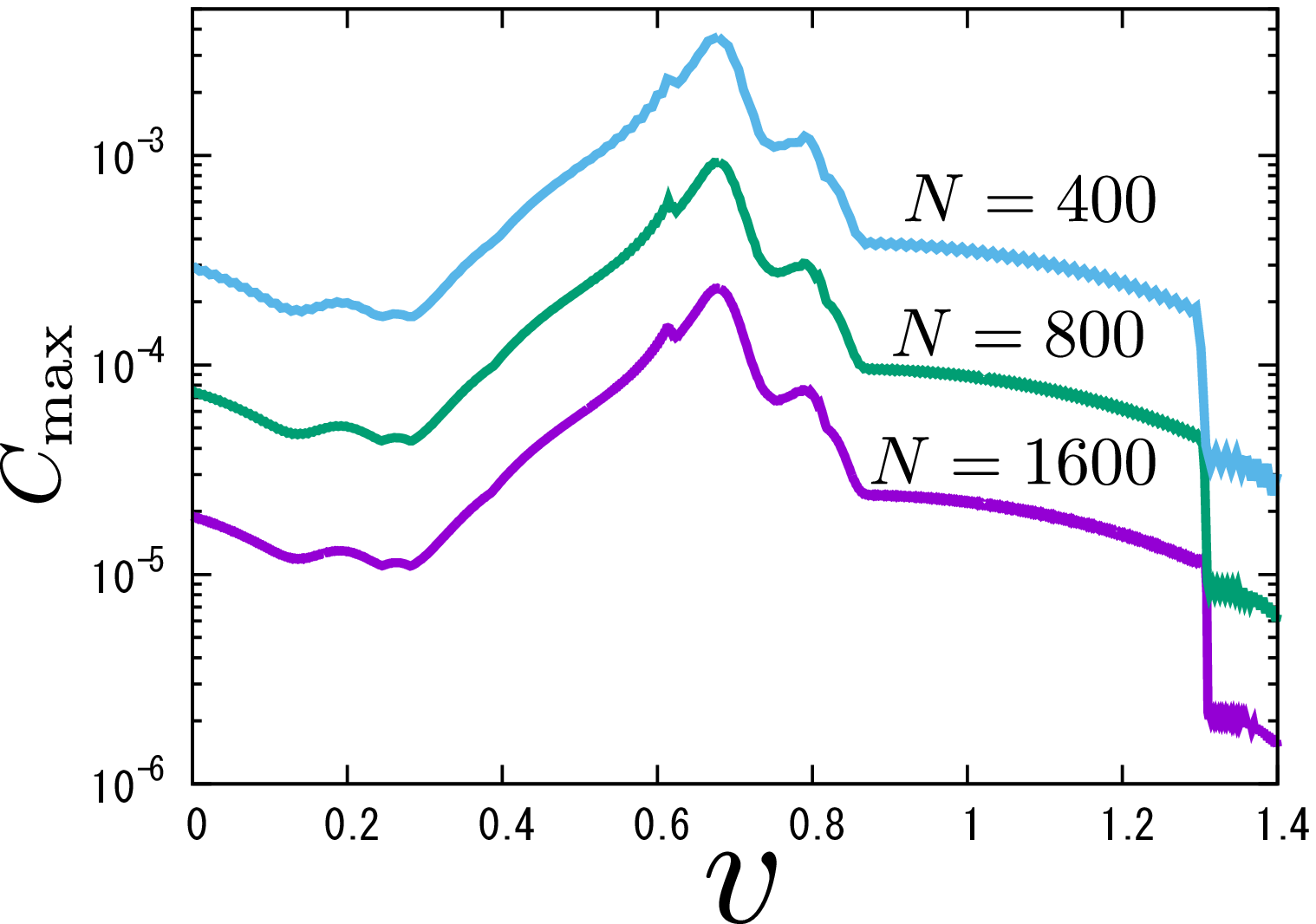}\label{CTC}
}
\caption{%
Constraint violation for several resolutions, $N=400$, $800$ and $1600$.
(a)~Longitudinal quench with $\epsilon=0.15$ and $\Delta t/L=2$.
(b)~Transverse linear quench with $\epsilon=0.15$ and $\Delta t/L=2$.
(c)~Transverse circular quench with $\epsilon=0.1$ and $\Delta t/L=2$.
\label{Cmax}
}
\end{figure}

For visibility, we consider the constraint violation on slices on the worldsheet.
We define a surface 
$\Sigma(v_0)\equiv \{v=v_0, \, 0<u<\beta_0, \, |\tau(u,v_0)|<\pi\}$. 
By assembling the surfaces, we define a one-dimensional function as
\begin{equation}
 C_\textrm{max}(v)=\max_{\Sigma(v)}(|\tilde{C}_1|,|\tilde{C}_2|)\ ,
\end{equation}
where on each $\Sigma(v)$ we seek the maximum value of the constraint violation by varying $u$.
We also take the bigger of $|\tilde{C}_1|$ and $|\tilde{C}_2|$.
Let $N$ be the number of the discretization segments on $v=\textrm{const}$ surfaces.
We plot $C_\textrm{max}(v)$ for $N=400$, $800$ and $1600$ in Fig.~\ref{Cmax}.
The constraint behaves as $C_\textrm{max}\propto 1/N^2$, consistent with 
our second-order discretization scheme.
In the paper, we basically choose $N=1600$ for which the constraint violation is less than $10^{-3}$.

\bibliography{bunkenF1G}

\end{document}